\documentclass[sigconf,nonacm]{acmart}

\ifdefined\pdfsuppressptexinfo
\fi
\hypersetup{pdfsource={},pdfauthor={Xiangwu Wang, Chengwei Cao, Hongyuan Tang}}

\usepackage{amsmath}
\usepackage{booktabs}
\usepackage{algorithm}
\usepackage[noend]{algpseudocode}
\usepackage{placeins}

\makeatletter
\@ACM@balancefalse
\makeatother

\setcopyright{none}
\renewcommand\footnotetextcopyrightpermission[1]{}
\acmConference{}{}{}
\acmDOI{}

\newtheorem{proposition}{Proposition}
\newtheorem{theorem}{Theorem}
\newtheorem{lemma}{Lemma}
\newcommand{\sirv}{\textsc{Sirv}}
\newcommand{\A}{\mathcal{A}}
\newcommand{\T}{\mathcal{T}}
\newcommand{\E}{\mathbb{E}}
\newcommand{\Prob}{\mathbb{P}}
\newcommand{\Gap}{\operatorname{Gap}}

\begin{document}

\title{Rank Without an Oracle: Deviation-Aware Interaction-Rank Selection from Offline Multi-Agent Logs}

\author{Xiangwu Wang}
\affiliation{\institution{University of Hong Kong}\country{}}

\author{Chengwei Cao}
\affiliation{\institution{University of California, San Diego}\country{}}

\author{Hongyuan Tang}
\affiliation{\institution{Carnegie Mellon University}\country{}}

\begin{abstract}
Offline multi-agent payoff models are estimated under a logging distribution
but used on distributions induced by learned solutions and unilateral
deviations.  Standard held-out loss can therefore favor an interaction class
that predicts logged play well while distorting strategic incentives.  We
introduce Selective Interaction-Rank Validation (\sirv) for finite games with
known logging distributions.  A training split fits nested payoff models and
constructs a common union of all candidate deployment and unilateral-
replacement distributions; an independent calibration split evaluates every
candidate on this same union.  \sirv{} returns the smallest rank whose
simultaneous upper worst-target risk is within tolerance of the best upper
score, and abstains when a declared target is unsupported or too imprecisely
estimated.  A common coverage event yields a finite-candidate target-risk
bound and a candidate-specific coarse correlated equilibrium (CCE) gap
certificate.  We also isolate an exact two-point off-support
non-identifiability result.  In a controlled factorial study with 2,048 independent games per
family, empirical-Bernstein bounds reduce the median CCE-gap certificate by
42.5\% relative to Hoeffding bounds on common returns, with a 1.36-point
reduction in supported return.  Under paired rank misspecification and in a
separately generated congestion family, the SIRV-EB fallback rule lowers mean
true candidate-selection CCE regret relative to ID-Mean, while retaining
game-level losses.  Across 384 games at $N=3,5,8$, ID-Mean-relative mean
CCE-regret effects stay positive while certified return falls sharply under weak coverage.
These results separate certifiable model selection from universal strategic
improvement.
\end{abstract}



\maketitle

\section{Introduction}

An offline multi-agent learner observes actions and payoffs under a fixed
logging distribution, yet its payoff model is used on distributions induced
by a learned solution and by agents deviating from that solution
~\cite{cui2022solvable,zhang2023markovgames}.  This shift makes model
complexity strategically consequential.  Low-interaction-rank models replace
unrestricted dependence on all agents with dependence on small interacting
groups and can improve statistical rates when the chosen rank captures the
relevant payoff structure~\cite{zhan2025exploiting}.  Existing analyses,
however, take the rank or function class as given.

Ordinary held-out error measures prediction under the logging distribution,
not whether a model preserves incentives under unilateral deviations.
Offline RL faces an analogous problem of ranking candidates without
deployment~\cite{tang2021modelselection}, while interaction-rank analyses take
the class as input and empirical-game analyses validate sampled-profile payoff
error~\cite{zhan2025exploiting,jordan2009generalization}.  Neither addresses
class selection on candidate-induced deviation distributions.
For example, under independent binary logging with action-one probability
$p$, a pair residual $ca_1a_2$ contributes $c^2p^2$ to logging risk but
$c^2p$ after fixing $a_1=1$, an amplification by $1/p$.  At $p=0$, that
off-support payoff cannot be recovered by reweighting.
Can interaction rank nevertheless be selected from logged play while every
candidate is compared on the deployment and deviation distributions induced
by the candidates themselves?

We address this question in finite normal-form games.  Selective
Interaction-Rank Validation (\sirv) first uses a training split to fit nested
models and a solution candidate for each rank, then defines the union of their
deployment and unilateral-replacement distributions.  Independent calibration
data provide simultaneous squared-risk intervals on this common union.  Thus
every rank faces the same targets.  \sirv{} abstains if the declared union is
unsupported or imprecise; otherwise it returns the smallest rank whose upper
score is within $\tau$ of the best upper score.

The common target union is the central object of our analysis.  It is fixed by
training data, yet supports a selective comparison on independent calibration
data.  On the simultaneous coverage event, the returned candidate's
worst-target risk is at most $\tau+\omega$ above that of the best candidate,
where $\omega$ is the observed maximum score width.  The same event yields a
candidate-specific coarse correlated equilibrium (CCE) gap certificate.
Because return is decided on calibration data, the bound and certificate
control the joint event of return and violation.  Complementing these
guarantees, a two-point construction identifies the information boundary:
without additional assumptions, deciding one specified off-support deviation
payoff has minimax error $1/2$.

We then evaluate whether target-risk calibration translates into strategic
performance, treating the two as separate empirical questions.  The study
combines 2,048-game factorial and misspecification panels with an independently
generated congestion family, target-scope ablations, and finite-grid scaling
through eight players.  True-payoff CCE regret, welfare regret, rank stability,
and certificate thresholds reveal both the gains from variance-adaptive bounds
and the remaining game-level losses.  Our scope is finite actions, known
logging probabilities, and targets fixed before calibration; the objective is
target-specific adequacy rather than algebraic-rank recovery.

\section{Related Work}

\subsection{Interaction Structure in Games}
Compact game representations exploit local or low-order interactions to avoid
the full joint-action description~\cite{kearns2001graphical}, including
context-specific, anonymous, and additive structure~\cite{jiang2011actiongraph}.
Empirical games have also used payoff regression, clustering, and graphical
structure recovery~\cite{vorobeychik2007learning,ficici2008learning,honorio2015learning}.
Zhan et al.~\cite{zhan2025exploiting} define interaction rank for offline
multi-agent reinforcement learning and show how low rank improves parts of the
distribution-shift and sample-complexity dependence.  Their guarantees
condition on a chosen $K$-IR class and assume that the relevant true and
learned rewards have interaction rank at most $K$; they do not select $K$ from
the log.  We instead select among nested ranks on candidate-induced targets,
rather than recover generating algebraic structure from partial payoff
observations~\cite{duong2009learning}.

\subsection{Empirical-Game Model Selection}
Jordan and Wellman~\cite{jordan2009generalization} compare representations of
empirical games using validation stratified over profiles, construct
reductions over equivalent strategies, and discuss factored representations;
strategic regret is evaluated after selection, not used as validation loss.
Related tools include importance-weighted cross-validation under covariate
shift~\cite{sugiyama2007iwcv} and overlap-aware confident off-policy
selection~\cite{kuzborskij2021confident}.

Single-agent offline RL provides oracle inequalities, selection without tuned
hyperparameters, uncertainty-aware or pessimistic comparisons, and
Bellman-error criteria~\cite{lee2022oracle,zhang2021hyperparameterfree,yang2022uncertainty,
yang2023pessimistic,zitovsky2023bellman}.  Logged-bandit counterfactual risk
minimization and overlap-aware evaluation likewise predate our estimators
~\cite{bottou2013counterfactual,swaminathan2015counterfactual,wang2017optimal}.

More broadly, data-dependent pessimism is a central response to insufficient
offline coverage, although its form determines the resulting bias--coverage
tradeoff~\cite{jin2021pessimism,xie2021bellman}.  Abstention itself has a long
history through reject-option and selective-prediction formulations, including
modern risk--coverage formulations and distribution-free risk control
~\cite{chow1970optimum,elyaniv2010foundations,bates2021risk}.
\sirv{} combines held-out, distribution-corrected validation with a common
multi-agent target set, an explicit support decision, and an equilibrium
certificate.  Here ``selective'' denotes an abstaining decision rule; the
guarantee controls the joint event of return and violation, not error
conditional on return.

\subsection{Offline-Game Coverage}
Offline-game theory has already established that coverage must account for
unilateral alternatives.  In two-player zero-sum Markov games, single-policy
coverage can be insufficient, whereas unilateral concentration supports
learning guarantees~\cite{cui2022solvable}.  Results for general function
approximation likewise control Nash, CE, or CCE learning through pessimism
and selective coverage of deviation policies~\cite{zhang2023markovgames}.
Strategy-wise confidence construction is another approach
~\cite{cui2022strategy}.  COPSRO and uncertainty-aware empirical-game solvers
choose equilibria under fixed representations
~\cite{shao2025copsro,nguyen2026conservative}.  Uniform payoff approximation
has also been related to equilibrium preservation and stable properties
~\cite{tuyls2020bounds,viqueira2020improved,cousins2023properties}; a recent
EGTA survey organizes the wider
representation, sampling, and equilibrium-analysis literature
~\cite{wellman2025egta}.  Taken together, prior empirical-game work selects
representations by predictive loss on sampled profiles, interaction-rank
analyses take the class as input, and offline-game methods select policies or
equilibria under a fixed representation.  \sirv{} instead selects the
representation class on one common, training-measurable union of rank-specific
deployment and unilateral-replacement distributions.  Relative to
uniform-payoff EGTA guarantees, its certificate is target-local to the returned
candidate.

\section{Problem Setting and Information Boundary}

Let $G=(N,(\A_i)_{i\in[N]},(u_i)_{i\in[N]})$ be a finite game,
$\A=\prod_i\A_i$, and $u_i:\A\to[0,1]$.  We observe i.i.d.\ tuples
$(A_s,Y_s)$, where $A_s\sim\nu$ for a known logging distribution $\nu$,
$Y_{s,i}\in[0,1]$, and $\E[Y_{s,i}\mid A_s=a]=u_i(a)$.  The data are split
once into training and calibration sets.  For binary actions, our nested rank
$k$ class contains Boolean monomials up to order $k$,
\begin{equation}
 f_{k,i}(a)=\sum_{S\subseteq[N],\,|S|\le k}\theta_{i,S}
 \prod_{j\in S}a_j , \qquad
 \mathcal F_1\subseteq\cdots\subseteq\mathcal F_{K_{\max}}.
 \label{eq:rankmodel}
\end{equation}
The method itself only needs nested fitted candidates; the theorem does not
require the true game to belong to one of these classes.
Our order counts total joint-action variables in a monomial.  For the
strategically relevant generator terms, which all contain player $i$, order
$k$ therefore means player $i$ plus at most $k-1$ other agents.  This makes
the indexing consistent with the $K$-IR convention of Zhan et
al.~\cite{zhan2025exploiting}, while Eq.~\eqref{eq:rankmodel} also permits
nuisance terms that do not contain $i$.

\subsection{Structural Rank and Target Adequacy}
Generating order need not equal the rank preferred for deployment: a
high-order term may be negligible on the declared targets, and estimation may
favor a smaller class.  Our evaluation oracle is therefore the smallest rank
whose exact worst risk on the training-defined target union is within $\tau$
of the minimum.  Synthetic payoff tables provide this label only for
evaluation and are unavailable to model fitting, target construction, or
selection.

To illustrate the distribution shift, consider an omission residual
$h(a)=c a_i a_j$ under product logging with action-one probabilities
$p_i,p_j$.  Its squared-risk contribution is $c^2p_ip_j$ under $\nu$ but
$c^2p_j$ under replacement $D_{i,1}(\nu)$, an amplification by
$1/p_i$ for $p_i>0$.  This calculation isolates the mass-shift mechanism; it
does not assume that a fitted model has exactly this residual.  Small $p_i$
causes imprecision, while $p_i=0$ creates the information boundary below.

For a distribution $q$ and pure replacement $b\in\A_i$, define
\begin{equation}
 D_{i,b}(q)(a)=\mathbf 1\{a_i=b\}
 \sum_{a_i'\in\A_i}q(a_i',a_{-i}).
 \label{eq:replacement}
\end{equation}
The CCE gap, including the no-deviation option, is
\begin{equation}
 \Gap_u(q)=\max\!\left\{0,\max_{i,b}
 \E_{A\sim q}[u_i(b,A_{-i})-u_i(A)]\right\}.
 \label{eq:ccegap}
\end{equation}
Pure replacements exhaust the ex-ante deviations in a finite CCE
definition~\cite{moulin1978strategically,monnot2017limits}.
CCE is also the canonical limiting notion associated with vanishing external
regret~\cite{roughgarden2015intrinsic}.

\subsection{Off-Support Non-Identifiability}
The following pair isolates one deviation payoff that the log cannot
distinguish.

\begin{proposition}[off-support deviation-payoff non-identifiability]
\label{prop:impossibility}
Consider three binary-action players, logging probabilities
$(0,0.25,0.5)$, and identical mean-zero additive noise supported on
$[-0.1,0.1]$.  Let all payoffs except player 1's be $0.5$, and let
\[
u^0_1(a)=0.4+0.1a_1,\qquad
u^1_1(a)=0.4+0.1a_1+0.2a_1a_2.
\]
Let $H\in\{0,1\}$ index the game and define the replacement payoff
\[
\begin{aligned}
\theta_H&=\E_{\nu_{-1}}[u^H_1(1,A_{-1})],
&(\theta_0,\theta_1)&=(0.50,0.55),\\[-2pt]
\inf_{\widehat H}\max_{h\in\{0,1\}}
\Pr_h(\widehat H\ne h)&=\frac12.
\end{aligned}
\]
Here the infimum ranges over possibly randomized tests based only on the log.
Equivalently, the binary decision target is
$H=\mathbf 1\{\theta_H>0.525\}$.
\end{proposition}

Logging fixes $a_1=0$, so the observation laws coincide, although the payoffs
are respectively 1-IR and 2-IR and differ by $0.05$ after replacement.
The proposition concerns this payoff point, not every partial-support
selection problem; Appendix~\ref{app:proof-impossibility} gives the proof.
A candidate-local rule could retain candidates whose own targets are
supported, but they would no longer be compared on the common union $\T$;
this changes $R_k$, the comparator, and the oracle.  \sirv{} instead uses a
conservative global support rule to retain the declared cross-candidate
target union.

\section{Selective Interaction-Rank Validation}
\label{sec:method}

Condition on the training split.  For every $k\in[K_{\max}]$, fit a payoff
model $\widehat u_k$ and compute a solver-selected CCE $\sigma_k$ of the
estimated game.  Every fitted payoff prediction is clipped to $[0,1]$.  We
deploy a logging-anchored mixture
\begin{equation}
q_k=(1-\alpha)\nu+\alpha\sigma_k,
\end{equation}
and define from the training split the common target set
\begin{equation}
\T=\bigcup_{j=1}^{K_{\max}}
\bigl(\{q_j\}\cup\{D_{i,b}(q_j):i\in[N],b\in\A_i\}\bigr).
\label{eq:targets}
\end{equation}
Every model is scored on every target in this union.  This prevents a rank
from being favored by generating an easy target only for itself.
It also accounts for cross-candidate deployment: rank $k$ must predict not
only $q_k$ but targets induced by every other rank.  The union is finite and
training-measurable, so this protection does not require uniform concentration
over an adaptive policy class.
The resulting objective is robust adequacy over the declared candidate set,
rather than candidate-local deployment-risk minimization.
Appendix~\ref{app:scaling-analysis} measures the return cost of this shared
comparator against candidate-local scopes, which define different risks and
oracles rather than competing estimates of the same objective.

For model $k$, player $i$, and target $t\in\T$, define noisy squared risk
\begin{equation}
r_{k,i,t}=\E_{A\sim t,Y_i\mid A}
[(\widehat u_{k,i}(A)-Y_i)^2],\qquad
R_k=\max_{i,t}r_{k,i,t}.
\label{eq:targetrisk}
\end{equation}
Conditional unbiasedness decomposes expected squared loss into squared payoff
bias plus reward variance.  Thus, under target support, $r_{k,i,t}$ is a
validation-risk estimand that upper-bounds mean payoff error.
Accordingly, $R_k$ is a predictive-risk estimand, not a structural-rank or
strategic-value estimand; we evaluate strategic outcomes separately.

Heterogeneous variance can still affect which player--target pair attains the
maximum; Appendix~\ref{app:limitations} discusses this scope.
The selection and certification steps accept any simultaneous confidence
procedure $\mathcal I$ that, conditional on training, returns
$[\ell_{k,i,t},u_{k,i,t}]$ satisfying the simultaneous coverage property
\begin{equation}
\Pr\!\left(\forall k,i,t:\ 
r_{k,i,t}\in[\ell_{k,i,t},u_{k,i,t}]
\mid D_{\rm tr}\right)\ge 1-\delta.
\label{eq:simultaneous-coverage}
\end{equation}
We satisfy this condition by estimating profile-conditional loss before
averaging under each target.  Let $n_a$ and $\widehat m_{k,i,a}$ be the
calibration count and squared-loss mean at profile $a$, and let
$F=K_{\max}N|\A|$.  The profile-Hoeffding construction uses, for $n_a>0$,
\begin{equation}
\rho^{\rm H}_a=\min\!\left\{1,
\sqrt{\frac{\log(2F/\delta)}{2n_a}}\right\}.
\label{eq:radius}
\end{equation}
The profile-EB construction instead uses the variance-adaptive
empirical-Bernstein radius derived in
Appendix~\ref{app:proof-oracle}.  Each construction uses $[0,1]$ when its required
profile count is unavailable, otherwise clips the profile interval to
$[0,1]$, and averages its endpoints with weights $t(a)$.  Both perform exact
finite-profile reweighting and satisfy
Eq.~\eqref{eq:simultaneous-coverage}~\cite{hoeffding1963probability,maurer2009empirical}.
Their relation to importance-weighted validation and overlap is
standard~\cite{sugiyama2007iwcv,cortes2010importance}.

The support decision uses the known population probabilities $\nu(a)$, while
the precision decision uses calibration counts and target mass.  A
positive-probability profile may therefore still trigger imprecision.  The
global support rule is not needed for interval validity---vacuous profile
intervals preserve coverage---but is a conservative policy that preserves the
declared target union.  Restricting $\T$ to supported targets is possible, but
would define a different risk, comparator, and oracle.

Define rank scores
\begin{equation}
L_k=\max_{i,t}\ell_{k,i,t},\quad
U_k=\max_{i,t}u_{k,i,t},\quad
\omega=\max_k(U_k-L_k).
\label{eq:scores}
\end{equation}
\sirv{} abstains if some target has positive mass outside the support of
$\nu$, or if the largest target-interval half-width exceeds a declared
$\epsilon_{\rm cert}$.  Otherwise it returns
\begin{equation}
\widehat k=\min\{k:U_k\le\min_jU_j+\tau\}.
\label{eq:selector}
\end{equation}
The tie-break returns the smallest rank meeting the declared upper-score
criterion.  The adequacy tolerance $\tau$ and precision threshold
$\epsilon_{\rm cert}$ are method parameters independent of the calibration
observations.  Algorithm~\ref{alg:sirv} summarizes
\sirv{};
full target-construction pseudocode appears in Appendix~\ref{app:algorithm}.

\begin{table}[t]
\caption{Selectors compared with the same split and tolerance.  Hats denote
calibration point estimates; H and EB denote Hoeffding and
empirical-Bernstein intervals.}
\label{tab:selectors}
\centering
\small
\begin{tabular}{@{}lll@{}}
\toprule
Method & Rank score & Refusal \\
\midrule
ID-Mean & mean player loss under $\nu$ & no \\
ID-Max & worst player loss under $\nu$ & no \\
Target-Point & $\max_{i,t}\widehat r_{k,i,t}$ & no \\
Target-IPS & $\max_{i,t}\widehat r^{\rm IPS}_{k,i,t}$ & support \\
IPS-EB & $\max_{i,t}u^{\rm IPS}_{k,i,t}$ & support/precision \\
\sirv-H & $\max_{i,t}u^{\rm H}_{k,i,t}$ & support/precision \\
\sirv-EB & $\max_{i,t}u^{\rm EB}_{k,i,t}$ & support/precision \\
\bottomrule
\end{tabular}
\end{table}

ID-Mean and ID-Max use average and worst-player logging loss, respectively;
Target-Point instead scores the common target set.  Target-IPS estimates those
risks by inverse propensity weighting, IPS-EB adds simultaneous confidence
bounds, and \sirv-H/\sirv-EB aggregate profile-level intervals.  Thus the
comparisons separate target choice, reweighting, and uncertainty treatment;
Table~\ref{tab:selectors} records their distinct refusal rules.

\begin{algorithm}[t]
\caption{Selective Interaction-Rank Validation (\sirv)}
\label{alg:sirv}
\begin{algorithmic}[1]
\Require split log, known $\nu$, ranks $1{:}K_{\max}$,
$\alpha,\tau,\epsilon_{\rm cert},\delta$, confidence procedure $\mathcal I$
\For{$k=1,\ldots,K_{\max}$}
  \State fit $\widehat u_k$ on training data; compute $\sigma_k$
  \State set $q_k\gets(1-\alpha)\nu+\alpha\sigma_k$
\EndFor
\State construct $\T$ from the training split using Eq.~\eqref{eq:targets}
\State build every $[\ell_{k,i,t},u_{k,i,t}]$ with $\mathcal I$
\If{a target is unsupported or maximum half-width $>\epsilon_{\rm cert}$}
  \State \Return \textsc{abstain}
\EndIf
\State compute $L_k,U_k$ and return Eq.~\eqref{eq:selector}
\end{algorithmic}
\end{algorithm}

\section{Guarantees}

Because all models and targets are measurable functions of the training
split, concentration can be applied conditionally on training.  The
profile-based confidence constructions in Section~\ref{sec:method} provide
the following event.

\begin{lemma}[simultaneous target intervals]
\label{lem:coverage}
The Hoeffding and empirical-Bernstein profile constructions, together with
the direct IPS-EB construction defined in Appendix~\ref{app:proof-oracle},
satisfy Eq.~\eqref{eq:simultaneous-coverage}.  Thus, with conditional probability at
least $1-\delta$, every $r_{k,i,t}$ lies in
$[\ell_{k,i,t},u_{k,i,t}]$ simultaneously.  Denote this event by
$E_{\rm target}$.
\end{lemma}

For either profile construction, a union bound first covers the profile risks; every
target then remains covered because it is a fixed nonnegative linear
combination of them.  IPS-EB instead covers target risks directly.  The result
does not assert usefulness: at zero or thin support an interval may be
vacuous.  All results below require only
$E_{\rm target}$ and therefore apply to any confidence procedure satisfying
Eq.~\eqref{eq:simultaneous-coverage}.

\begin{theorem}[finite-candidate target-risk selection bound]
\label{thm:oracle}
Let $\mathcal I$ be any confidence procedure satisfying
Eq.~\eqref{eq:simultaneous-coverage}, and let
$k^*\in\arg\min_kR_k$.  On $E_{\rm target}$, whenever \sirv{} returns,
\begin{equation}
R_{\widehat k}\le \min_kR_k+\tau+\omega.
\label{eq:oraclebound}
\end{equation}
Consequently,
\begin{equation}
\Prob\!\left(\text{\sirv{} returns and }
R_{\widehat k}-\min_kR_k>\tau+\omega\right)\le\delta.
\label{eq:jointoracle}
\end{equation}
\end{theorem}

The proof is a short interval chain:
$R_{\widehat k}\le U_{\widehat k}\le U_{k^*}+\tau
\le L_{k^*}+\omega+\tau\le R_{k^*}+\omega+\tau$.
Notably, the selector compares upper scores to the best upper score; it does
not compare against the best lower bound.  The final probability statement is
joint.  Conditioning it on return would divide by a data-dependent return
probability and is not justified by this argument.

\begin{proposition}[observable precision and shared failure event]
\label{prop:precision}
If \sirv{} returns, then $\omega\le2\epsilon_{\rm cert}$.  Moreover, the
probability that it returns and violates either Eq.~\eqref{eq:oraclebound} or
the CCE certificate in Theorem~\ref{thm:cce} is at most $\delta$ (not
$2\delta$).
\end{proposition}

For each rank, let $(i,t)$ attain $U_k$.  Then
$U_k-L_k\le u_{k,i,t}-\ell_{k,i,t}\le2\epsilon_{\rm cert}$ on return, proving
the first statement.  Both theorem violations are impossible on the same
event $E_{\rm target}$, so their union is contained in
$E_{\rm target}^{c}$.  The precision term is therefore observable before
deployment, although conservatism in the component intervals can make it
larger than the realized excess risk.

Target risk matters strategically because CCE gain is a difference of two
payoff expectations.  Define the computable certificate
\begin{equation}
C_k=\Gap_{\widehat u_k}(q_k)
+\max_i\sqrt{u_{k,i,q_k}}
+\max_{i,b}\sqrt{u_{k,i,D_{i,b}(q_k)}}.
\label{eq:gapcert}
\end{equation}

\begin{theorem}[candidate-specific CCE certificate]
\label{thm:cce}
On $E_{\rm target}$, $\Gap_u(q_k)\le C_k$ for every candidate $k$.
Therefore
\begin{equation}
\Prob(\text{\sirv{} returns and }\Gap_u(q_{\widehat k})>C_{\widehat k})
\le\delta.
\label{eq:jointgap}
\end{equation}
\end{theorem}

For each deviation, insert and subtract the predicted payoff difference.  Its
two bias terms are bounded by Cauchy--Schwarz under $q_k$ and
$D_{i,b}(q_k)$.  Since $Y_i$ is conditionally unbiased,
$\E[(\widehat u_{k,i}-Y_i)^2\mid A]$ upper-bounds squared mean-payoff error.
Maximizing gives Eq.~\eqref{eq:gapcert}; Appendix~\ref{app:proof-cce} gives the
full proof.  This theorem certifies a returned candidate.  It neither orders
candidates by true CCE gap nor promises that \sirv{} improves that gap over
in-distribution validation.

The predicted-gap term in Eq.~\eqref{eq:gapcert} is necessary.  Although
$\sigma_k$ is a CCE of the estimated game, the deployed
$q_k=(1-\alpha)\nu+\alpha\sigma_k$ need not be one.  The certificate therefore
does not assume that logging anchoring preserves equilibrium.  Also,
because $r_{k,i,t}$ includes conditional reward variance, it upper-bounds
mean-payoff error rather than estimating it exactly; this is one source of the
certificate slack observed below.

\section{Experiments}

We evaluate SIRV through strategic outcomes and certificate utility computed
from the true payoff table, neither of which defines its selection objective.
We then study robustness to rank misspecification and heterogeneous noise.
Target-risk alignment is reported separately in
Appendix~\ref{app:additional}.

\subsection{Design}

We generate three-player binary-action games with $z_j=2a_j-1$ and
\begin{equation}
u_i(a)=0.5+\sum_{1\le |S|\le K,\,i\in S}
\theta_{i,S}\prod_{j\in S}z_j .
\label{eq:generator}
\end{equation}
The controlled family makes target risks, CCE gaps, and welfare exactly
computable, separating certificate validity from outcome-estimation error.
Within each interaction order, independent signs and random magnitudes are
normalized to fixed $\ell_1$ budgets.  Rank 1 uses $B_1=0.16$; rank 2 uses
$(B_1,B_2)=(0.16,h)$; and rank 3 uses $(0.08,0.08,h)$, with
$h\in\{0.04,0.12\}$.  The total budget keeps noiseless payoffs in
$[0.22,0.78]$, and observations add independent
$\operatorname{Unif}[-0.02,0.02]$ noise.

Each order contains 2,048 independent base games.  Product-Bernoulli
logging is balanced, moderate, thin, or unsupported; the three supported
regimes have minority-action probabilities $0.5$, $[0.25,0.40]$, and
$[0.08,0.18]$, respectively.  Unsupported logging fixes one action
coordinate.  We cross these regimes with
$n\in\{20{,}000,80{,}000,320{,}000\}$ and two repetitions.  The five game
families therefore yield 245,760 configurations, of which 184,320 are supported.
Conditions within a base game share the underlying payoff coefficients, and
the base game is the inferential unit.

All selectors use the same 50/50 training--calibration split, candidate ranks
$\{1,2,3\}$, solver-selected welfare-maximizing predicted CCEs, and
deployment weight $\alpha=0.5$.  We set $\tau=0.002$,
$\epsilon_{\rm cert}=0.12$, and $\delta=0.05$.  The two in-distribution
comparators are mean logging loss (ID-Mean) and worst-player logging loss
(ID-Max).  Target-Point evaluates point estimates on the common target union.
The same parameter setting is used throughout.
Target-IPS instead estimates each target risk directly as
\[
\widehat r^{\rm IPS}_{k,i,t}
=\frac{1}{m}\sum_{s=1}^{m}
\frac{t(A_s)}{\nu(A_s)}
(\widehat u_{k,i}(A_s)-Y_{s,i})^2 .
\]
The seven selectors in Table~\ref{tab:selectors} share data, candidate
deployments, target union, and tolerance, but differ in support and precision
behavior.

For selected rank $\widehat k$, we measure candidate CCE regret and welfare
regret using the simulator's mean payoff table, with
$W_u(q)=\E_{A\sim q}[N^{-1}\sum_i u_i(A)]$:
\begin{align}
\operatorname{Reg}_{\rm CCE}(\widehat k)
&=\Gap_u(q_{\widehat k})-\min_{k\in\mathcal K}\Gap_u(q_k),\\
\operatorname{Reg}_{W}(\widehat k)
&=\max_{k\in\mathcal K} W_u(q_k)-W_u(q_{\widehat k}).
\end{align}
These candidate-selection regrets compare only the fitted deployments
$\{q_k\}$, not an unrestricted strategic oracle, and are unavailable to every
selector.  For selective methods, we report both conditional-on-return
quantities and a complete-decision rule that uses ID-Mean after abstention.
The main paired comparisons use this ID-fallback rule over all supported rows,
with positive effects favoring the alternative to ID-Mean.
We average paired effects within each base game and report equal-game means
and 95\% percentile intervals from 5,000 bootstrap resamples of the 2,048
games~\cite{wiedenbeck2014bootstrap}.  Intervals quantify random-game
variation conditional on the fixed grid; within-game conditions are averaged,
not treated as independent units.
The primary comparison is SIRV-EB
versus ID-Mean for candidate CCE regret under heterogeneous noise; other
intervals are descriptive, with ties and losses retained.

The robustness study applies a paired perturbation to the same 2,048 rank-3,
$h=0.12$ base games used in the factorial study; it is not an independent
holdout set.  Every selector is restricted to ranks $\{1,2\}$, so neither
candidate is correctly specified.  It
crosses balanced, moderate, and thin logging with the same three sample sizes
and two repetitions.  Under homoscedastic noise, observations again use
$\operatorname{Unif}[-0.02,0.02]$.  Under heterogeneous noise, each
game--player--profile has a half-width $b_{i,a}\in[0.01,0.10]$ and noise
$\operatorname{Unif}[-b_{i,a},b_{i,a}]$; exact target risk uses variance
$b_{i,a}^2/3$.  The two noise conditions contribute 36,864 rows each, for
73,728 configurations.  True CCE and welfare outcomes use only the mean payoff
table.  Regret in this study is relative to the better deployment
within $\mathcal K=\{1,2\}$, not to the excluded rank-3 model or an
unrestricted oracle.

\subsection{Selective return}

\begin{figure*}[!t]
  \centering
  \includegraphics[width=.98\textwidth]{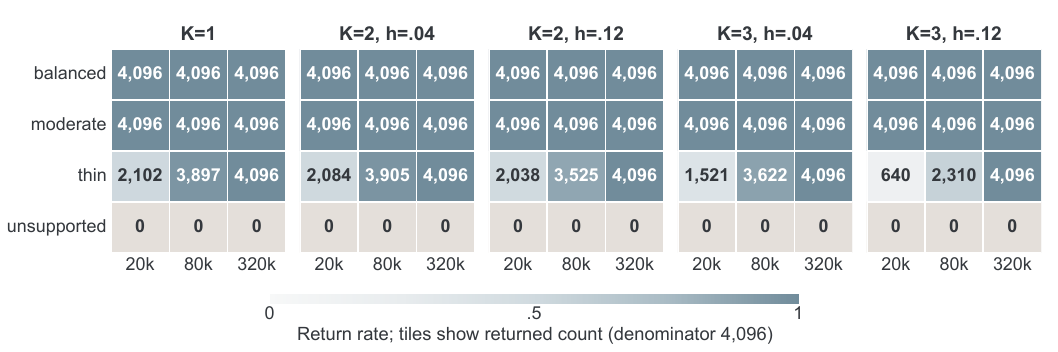}
  \caption{SIRV-EB selective return over the complete factorial grid.
  Five game families, four logging regimes, and three sample sizes give 60
  cells.  Each tile reports returns out of 4,096 runs (2,048 base games and two
  repetitions), accounting for all 245,760 configurations.  Return increases
  with sample size under thin logging; all unsupported cells are refused.}
  \Description{Five side-by-side four-by-three tile matrices show SIRV-EB
  returned-run counts.  Rows are balanced, moderate, thin, and unsupported
  logging; columns are 20k, 80k, and 320k samples.  Thin-row counts increase
  with sample size, while every unsupported count is zero.}
  \label{fig:return}
\end{figure*}

SIRV-EB returned on 169,004/184,320 supported rows (91.69\%) and refused all
61,440 unsupported rows (Figure~\ref{fig:return}).  The SIRV-H control returned
on 171,504 supported rows (93.05\%).  Direct IPS-EB returned on 166,611
(90.39\%).  Target-IPS returned on every supported row and refused every
unsupported row.  Counts by condition and repetition stability appear in
Appendix~\ref{app:additional}.

\subsection{Strategic outcomes in the well-specified family}

\begin{figure*}[!t]
  \centering
  \includegraphics[width=.98\textwidth]{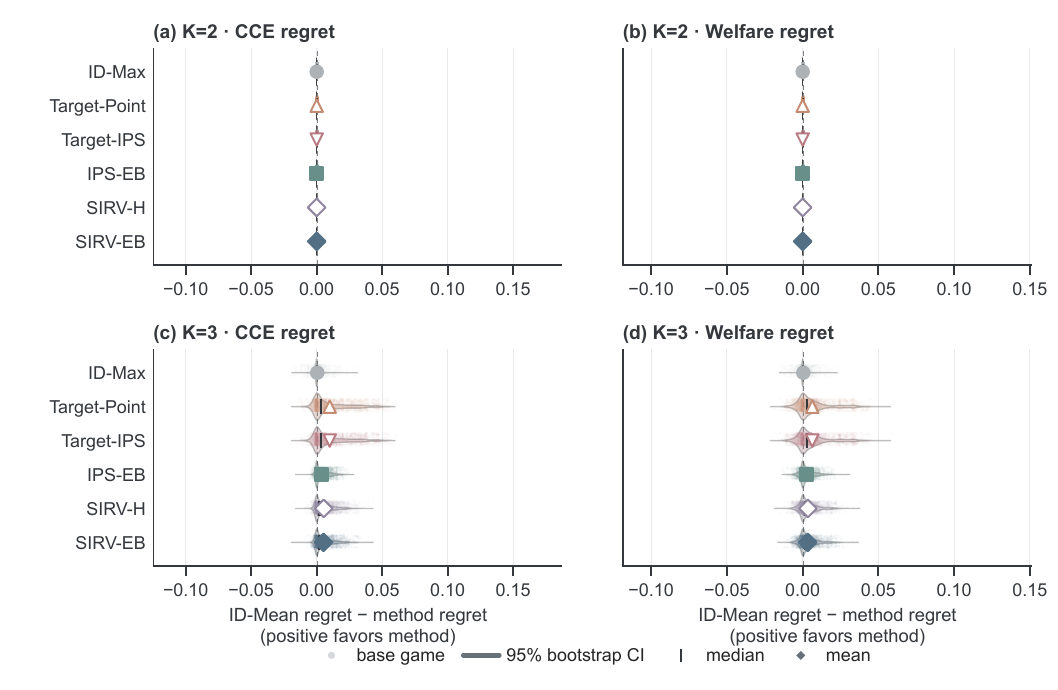}
  \caption{External strategic effects in the well-specified study.  Rows
  separate generating ranks 2 and 3; columns show true-payoff
  candidate-selection CCE regret
  and welfare regret.  Each panel contains the 2,048 paired base-game effects for six
  alternatives to ID-Mean under ID fallback, with interval summaries computed
  over games.  Positive values favor the alternative.  Every rank-2 effect is
  an exact tie; rank 3 has positive means but retains method-dependent ties
  and negative game-level effects.}
  \Description{Four paired-effect panels compare ID-Max, Target-Point,
  Target-IPS, IPS-EB, SIRV-H, and SIRV-EB with ID-Mean.  Both rank-two panels
  lie at zero.  Rank-three CCE and welfare panels show positive mean effects
  with distributions extending to zero or negative values for some methods.}
  \label{fig:external}
\end{figure*}

The external outcomes sharply separate generating orders 2 and 3
(Figure~\ref{fig:external}).  Generating-order-1 and -2 families each tied
ID-Mean exactly in both regrets across all 2,048 games; order 1 is omitted
from the figure.  For generating order 3,
the SIRV-EB fallback
system reduced mean CCE regret by 0.00518 (95\% interval
$[0.00486,0.00550]$) and welfare regret by 0.00328
($[0.00303,0.00355]$).  SIRV-H produced similar effects, 0.00538 and
0.00346.  The point selectors had larger mean effects:
Target-Point and Target-IPS each reduced CCE regret by 0.00993 and welfare
regret by 0.00635.  They do not, however, supply the simultaneous selective
certificates evaluated below.  Thus the results distinguish strategic mean
performance from certifiability rather than defining a single method ranking.

\subsection{Rank misspecification and heterogeneous noise}

\begin{figure*}[!t]
  \centering
  \includegraphics[width=\textwidth]{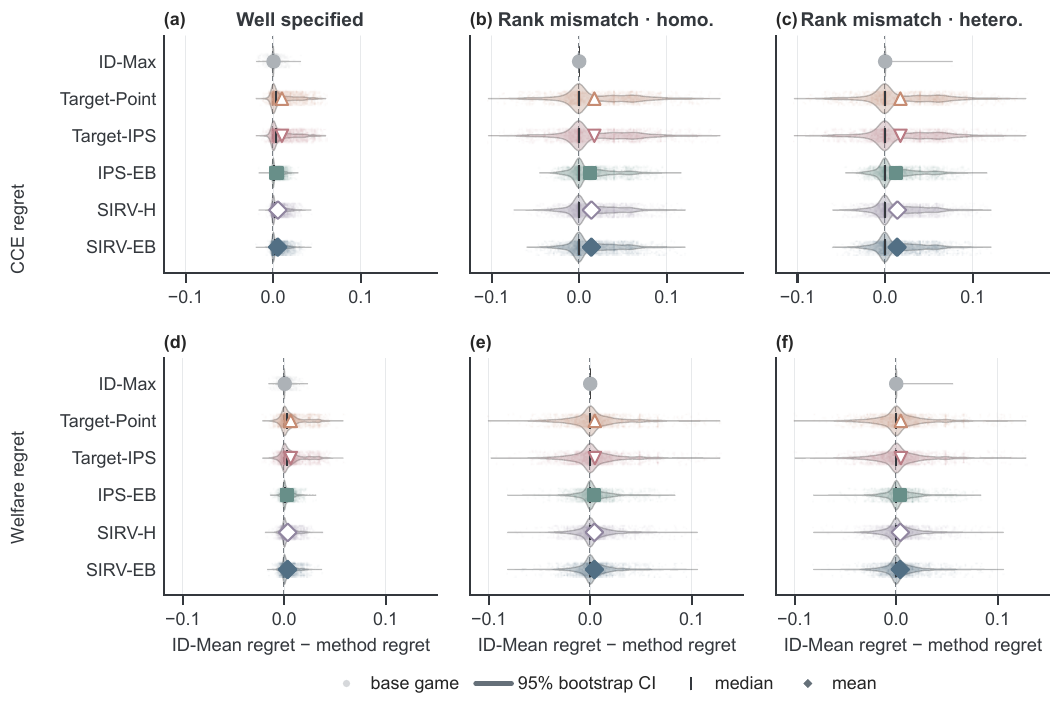}
  \caption{External effects across specification and noise conditions.
  Rows show true-payoff candidate-selection CCE and welfare regret; columns show well-specified
  rank-3 games, rank-misspecified games with homoscedastic noise, and the same
  misspecified games with heterogeneous noise.  Each panel contains six
  methods' 2,048 paired game effects relative to ID-Mean under ID fallback.
  Positive values favor the alternative.}
  \Description{Six panels compare paired game-level regret effects.  CCE
  effects have positive means for target-aware methods in all three columns,
  but include losses and ties.  Target-aware welfare effects are positive on
  average in the three columns, while their game-level distributions also
  include losses and ties.}
  \label{fig:misspecified}
\end{figure*}

Under misspecification and heterogeneous noise, the SIRV-EB fallback rule
reduced mean candidate CCE regret by 0.01362 relative to ID-Mean
(95\% interval $[0.01248,0.01479]$; Figure~\ref{fig:misspecified}).  The effect was
positive for 49.2\% of game means, tied for 25.0\%, and negative for 25.9\%.
The corresponding mean welfare-regret effect was 0.00397, with interval
$[0.00321,0.00474]$; 32.1\% of its game-level effects were negative.
Homoscedastic noise gave a similar SIRV-EB CCE effect,
0.01383 $[0.01265,0.01502]$, and a mean welfare effect of 0.00402
$[0.00324,0.00479]$.

Target-Point attained a larger mean CCE effect under
heterogeneous noise: 0.01741, with a 95\% interval from 0.01569 to 0.01915.
The Target-IPS mean was 0.01746, with interval 0.01576 to 0.01917.  Neither
point-estimator result removes the negative game-level tail, and neither
establishes pointwise dominance.
These results support positive average CCE- and welfare-regret effects in this
controlled misspecification family, not a universal strategic advantage.
On 512 independent weighted congestion games, SIRV-EB fallback improved mean
CCE/welfare regret by 0.01317/0.01608, with both 95\% game-bootstrap intervals
above zero (Appendix~\ref{app:congestion}).
On the $N\in\{3,5,8\}$ scale grid, its mean CCE effect remained positive,
while return and certificate utility declined; three of nine
agent-count--logging cells did not meet the predefined criterion through 1.28
million samples (Appendix~\ref{app:scaling-analysis}).

\subsection{Certificate utility}

\begin{figure*}[!t]
  \centering
  \includegraphics[width=\textwidth]{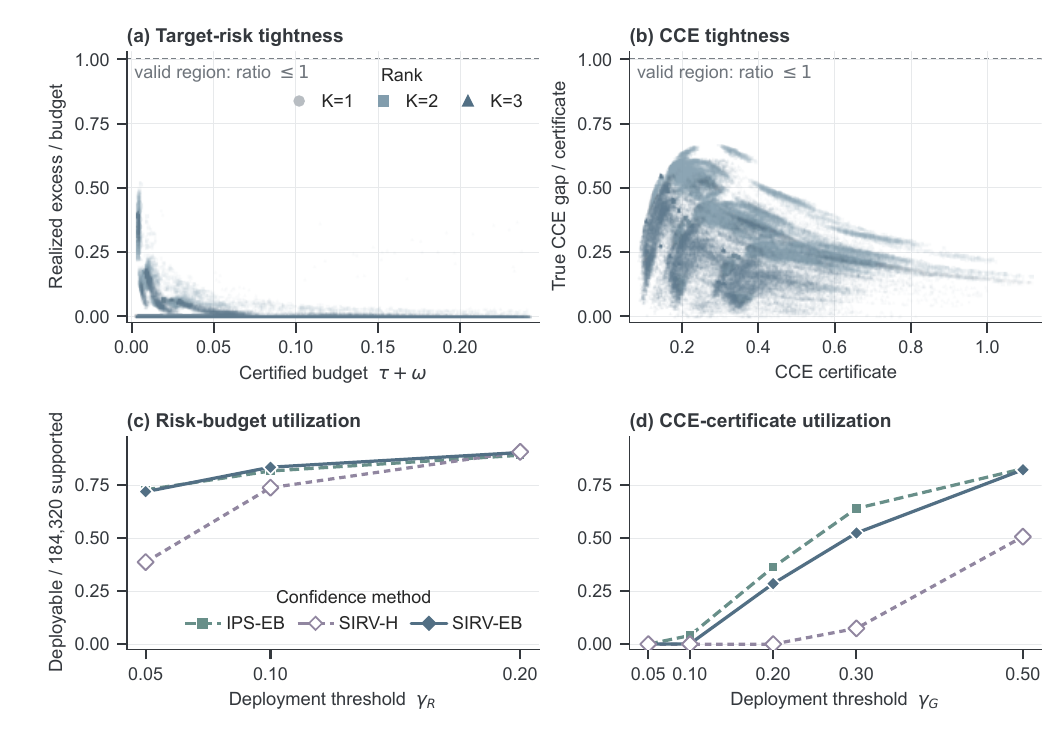}
  \caption{Certificate tightness and threshold utility on the 184,320 supported
  well-specified rows.  Panels (a--b) show all 169,004 SIRV-EB returns:
  realized target-risk excess divided by budget $\tau+\omega$, and true CCE gap
  divided by its candidate certificate, plotted against the corresponding
  certificate quantity.
  Panels (c--d) report the fraction of all supported rows that both return and
  satisfy each deployment threshold for SIRV-H, SIRV-EB, and IPS-EB;
  abstentions remain in the denominator.}
  \Description{The upper panels compare realized quantities with their
  certificate boundaries for SIRV-EB returns.  The lower panels show
  threshold-response profiles for three confidence procedures.  Empirical-
  Bernstein bounds certify substantially more rows at strict risk and CCE
  thresholds than SIRV-H, while the strictest CCE thresholds remain rarely or
  never attainable.}
  \label{fig:certificate}
\end{figure*}

On 168,904 common returns, the median CCE certificate was 0.26668 for SIRV-EB
versus 0.46365 for SIRV-H, a 42.5\% reduction.  Overall supported return was
1.36 percentage points lower.  Across its own returns, IPS-EB reached a
median of 0.23296 with a 90.39\% return rate.  Among all supported rows, the
fraction that both returned and had risk
budget at most 0.05 rose from 38.73\% for SIRV-H to 71.94\% for SIRV-EB and
72.83\% for IPS-EB.  At a CCE threshold of 0.2, the corresponding fractions
were 0\%, 28.57\%, and 36.34\%; at 0.5 they were 50.69\%, 82.29\%, and
82.80\% (Figure~\ref{fig:certificate}).  No method produced a certificate at
most 0.05, so the certificates remain conservative at strict strategic
tolerances.

Among SIRV-EB returns, the median realized-excess/budget ratio was zero and
its 90th percentile was 0.161; the median true-gap/certificate ratio was
0.282.  The heterogeneous-noise condition was harder to certify: its median
SIRV-EB CCE certificate was 0.40015, and only 0.65\%, 7.69\%, and 77.63\% of
returns met CCE thresholds 0.2, 0.3, and 0.5.  Requiring both
$\tau+\omega\le0.05$ and $C_{\widehat k}\le0.2$ retained 28.21\% of all
supported factorial rows but only 0.56\% of heterogeneous-noise rows.
Every covered return in both studies satisfied its target-risk budget and CCE
certificate.  A user with
acceptable risk excess $\gamma_R$ can require
$\tau+\omega\le\gamma_R$, and a user with acceptable strategic gap
$\gamma_G$ can additionally require $C_{\widehat k}\le\gamma_G$.  The
threshold profiles show the resulting utilization rather than reporting
coverage alone.

\begin{table}[H]
\caption{Selected-rank agreement across the two repetitions under ID
fallback.  The factorial and misspecified heterogeneous-noise columns contain 92,160 and
18,432 supported repetition pairs, respectively.}
\label{tab:stability}
\centering
\small
\begin{tabular}{lrr}
\toprule
Method & Factorial & Misspecified hetero. \\
\midrule
ID-Mean      & .9959 & .9989 \\
ID-Max       & .9933 & .9995 \\
Target-Point & .9978 & .9823 \\
Target-IPS   & .9972 & .9700 \\
IPS-EB       & .9922 & .9852 \\
SIRV-H       & .9896 & .9833 \\
SIRV-EB      & .9892 & .9826 \\
\bottomrule
\end{tabular}
\end{table}

Under ID fallback, repetition agreement was at least .9892 in the factorial
study and .9700 in the misspecified study.  Return-decision agreement ranged
from .9915 to .9974 and from .9869 to .9909, respectively.
\FloatBarrier

\section{Conclusion}

Offline multi-agent model selection should evaluate candidates on deployment
and unilateral deviations, not only logged play.  \sirv{} compares nested
ranks on a common target union, returning the smallest rank that meets a
simultaneous upper-score criterion or abstaining.  One coverage event yields
target-risk and candidate-specific CCE-gap bounds; a two-point construction
establishes non-identifiability of a specified off-support payoff.

Empirical-Bernstein intervals reduced median CCE certificates by 42.5\%
relative to Hoeffding intervals.  The SIRV-EB fallback system had positive
average CCE- and welfare-regret effects in the rank-3, misspecified, and
congestion families, while rank-2 effects were zero and individual losses
remained.  Scaling retained positive mean effects as certificate utility
declined.  Thus \sirv{} provides auditable, target-aware selection rather
than universal strategic improvement; its return decision can be combined
with application-specific risk and CCE thresholds.

\label{page:last-main}
\clearpage
\label{page:first-reference}
\setlength{\bibsep}{0pt plus 0.3ex}
\bibliographystyle{ACM-Reference-Format}
\bibliography{references}

\clearpage
\appendix
\section{Notation and Assumptions}
\label{app:notation}

Table~\ref{tab:notation} collects the quantities used in the proofs.  The
minimal statistical assumptions are: (i) calibration observations are
independent conditional on the training split; (ii) $Y_i\in[0,1]$ and
$\E[Y_i\mid A=a]=u_i(a)$; (iii) fitted predictions are in $[0,1]$; (iv) the
finite models, candidates, deployments, and targets are measurable functions
of training data only; and (v) the logging distribution $\nu$ is known for the
support check.  Realizability is not assumed for the risk or CCE theorem.

\begin{table}[ht]
\caption{Core notation.}
\label{tab:notation}
\centering
\small
\begin{tabular}{ll}
\toprule
Symbol & Meaning \\
\midrule
$N,\A$ & players and finite joint-action set \\
$\nu$ & known logging distribution \\
$\widehat u_k$ & rank-$k$ fitted payoff vector \\
$\sigma_k$ & solver-selected CCE of $\widehat u_k$ \\
$q_k$ & $(1-\alpha)\nu+\alpha\sigma_k$ \\
$D_{i,b}(q)$ & pure replacement of player $i$ by action $b$ \\
$\T$ & common training-defined target union \\
$m_{k,i,a}$ & conditional noisy squared loss at profile $a$ \\
$r_{k,i,t}$ & target-weighted noisy squared risk \\
$R_k$ & worst player/target risk of rank $k$ \\
$[L_k,U_k]$ & max-aggregated rank score bounds \\
$\omega$ & $\max_k(U_k-L_k)$ \\
$E_{\rm target}$ & simultaneous target-interval event \\
\bottomrule
\end{tabular}
\end{table}

\section{Proofs for the Confidence Bounds and Selection Guarantee}
\label{app:proof-oracle}

\subsection{Proof of Lemma~\ref{lem:coverage}}

Condition on all training data.  This fixes $\widehat u_k$, $\sigma_k$, $q_k$,
and $\T$.  For a joint profile $a$, the calibration losses
\[
Z_s=(\widehat u_{k,i}(a)-Y_{s,i})^2\in[0,1]
\]
among observations satisfying $A_s=a$ are independent with mean
$m_{k,i,a}$.  Let $F=K_{\max}N|\A|$.

For the Hoeffding construction and $n_a>0$, the two-sided inequality gives
\[
\Pr\!\left(
|\widehat m_{k,i,a}-m_{k,i,a}|>\rho^{\rm H}_a
\mid D_{\rm tr},n_a\right)
\le 2e^{-2n_a(\rho^{\rm H}_a)^2}\le\frac{\delta}{F}.
\]
If the radius clips at one, or if $n_a=0$, the clipped interval $[0,1]$
is valid deterministically.

For the empirical-Bernstein construction and $n_a\ge2$, let
$\widehat v_{k,i,a}$ be the unbiased sample variance of these losses.
It uses
\begin{equation}
\rho^{\rm EB}_{k,i,a}=\min\!\left\{1,
\sqrt{\frac{2\widehat v_{k,i,a}\log(4F/\delta)}{n_a}}
+\frac{7\log(4F/\delta)}{3(n_a-1)}\right\}.
\label{eq:ebradius}
\end{equation}
Applying the empirical-Bernstein inequality to both tails gives
\[
\Pr\!\left(
|\widehat m_{k,i,a}-m_{k,i,a}|>\rho^{\rm EB}_{k,i,a}
\mid D_{\rm tr},n_a\right)\le\frac{\delta}{F},
\]
for the radius in Eq.~\eqref{eq:ebradius}.  When $n_a<2$, the construction instead
uses $[0,1]$.  Thus a union bound over all $F$ triples $(k,i,a)$ shows that
either construction covers every profile risk with conditional probability at least
$1-\delta$.

For a target $t$ determined from the training split,
\[
r_{k,i,t}=\sum_{a\in\A}t(a)m_{k,i,a}.
\]
The weights are nonnegative and sum to one.  Therefore simultaneous profile
coverage implies
\[
\sum_a t(a)\ell_{k,i,a}\le r_{k,i,t}
\le\sum_a t(a)u_{k,i,a},
\]
which is exactly the constructed target interval.  Averaging the conditional
failure probability over training proves the unconditional statement for
both profile constructions.

\paragraph{Direct IPS-EB.}
For a supported target $t$, let
$w_t(a)=t(a)/\nu(a)$,
$B_t=\max_{a:\nu(a)>0}w_t(a)$, and
\[
X_s=w_t(A_s)
(\widehat u_{k,i}(A_s)-Y_{s,i})^2 .
\]
Conditional on training, $\E X_s=r_{k,i,t}$ and
$X_s\in[0,B_t]$.  With candidate set $\mathcal K$, define
$F_{\rm IPS}=|\mathcal K|N|\T|$.  For calibration size $m\ge2$ and unbiased
sample variance $\widehat v^{\rm IPS}_{k,i,t}$, IPS-EB uses
\begin{equation}
\rho^{\rm IPS}_{k,i,t}
=\sqrt{\frac{2\widehat v^{\rm IPS}_{k,i,t}
\log(4F_{\rm IPS}/\delta)}{m}}
+\frac{7B_t\log(4F_{\rm IPS}/\delta)}{3(m-1)}.
\label{eq:ips-eb-radius}
\end{equation}
Applying the empirical-Bernstein inequality to $X_s/B_t$ and
$(B_t-X_s)/B_t$, followed by a union bound over $(k,i,t)$, gives simultaneous
two-sided coverage~\cite{maurer2009empirical}.  Unsupported targets or
$m<2$ receive $[0,1]$; otherwise the interval is
\[
\left[
\widehat r^{\rm IPS}_{k,i,t}-\rho^{\rm IPS}_{k,i,t},\
\widehat r^{\rm IPS}_{k,i,t}+\rho^{\rm IPS}_{k,i,t}
\right]\cap[0,1].
\]
Thus IPS-EB also satisfies Eq.~\eqref{eq:simultaneous-coverage}.  Averaging over
training completes the proof of Lemma~\ref{lem:coverage}.
\hfill$\square$

\subsection{Proof of Theorem~\ref{thm:oracle}}

On $E_{\rm target}$, max aggregation preserves every score interval:
$L_k\le R_k\le U_k$.  Let $k^*$ minimize $R_k$.  If \sirv{} returns, its rule
and the definition of $\omega$ yield
\begin{align*}
R_{\widehat k}
&\le U_{\widehat k} \\
&\le \min_j U_j+\tau \\
&\le U_{k^*}+\tau \\
&\le L_{k^*}+\omega+\tau \\
&\le R_{k^*}+\omega+\tau.
\end{align*}
This proves Eq.~\eqref{eq:oraclebound}.  A return-and-violation can occur only
on $E_{\rm target}^c$, so its probability is at most $\delta$.
\hfill$\square$

The exact evaluation oracle used in the experiments is
\[
k^*(\tau)=\min\{k:R_k\le\min_j R_j+\tau\}.
\]
It is an evaluation label, not an input to the theorem or selector.  The
finite-candidate bound compares the selected risk with the minimum exact risk;
accuracy measures whether the selected rank equals this smallest exact
target-risk-oracle rank.  These are related but distinct evaluations.

\subsection{Joint and conditional violation probabilities}

Let $B$ be either violation event.  The proof gives
$\Prob(\mathrm{return}\cap B)\le\delta$.  In contrast,
\[
\Prob(B\mid\mathrm{return})
=\frac{\Prob(\mathrm{return}\cap B)}{\Prob(\mathrm{return})}
\le\frac{\delta}{\Prob(\mathrm{return})},
\]
when the denominator is positive.  Because the precision criterion is calculated
from calibration data, return is not independent of interval coverage.
Accordingly, the theorem controls the joint event, and empirical system-level
denominators retain abstentions.

\section{Proof of the CCE-Gap Certificate}
\label{app:proof-cce}

Fix candidate $k$, player $i$, and replacement $b$.  Write $f=\widehat
u_{k,i}$, $q=q_k$, and $d=D_{i,b}(q)$.  By construction of $d$,
\[
\E_{A\sim q}u_i(b,A_{-i})=\E_{A\sim d}u_i(A),
\]
and the same identity holds for $f$.  Insert and subtract the predicted
payoffs:
\begin{align*}
&\E_q[u_i(b,A_{-i})-u_i(A)]\\
&=\E_q[f(b,A_{-i})-f(A)]
 +\E_d[u_i-f]-\E_q[u_i-f]\\
&\le \E_q[f(b,A_{-i})-f(A)]
 +|\E_d[u_i-f]|+|\E_q[u_i-f]|.
\end{align*}
Cauchy--Schwarz gives
\[
|\E_t[u_i-f]|\le\sqrt{\E_t[(u_i-f)^2]}
\quad\text{for }t\in\{q,d\}.
\]
Conditional unbiasedness supplies the variance decomposition
\[
\E[(f(A)-Y_i)^2\mid A=a]
=(f(a)-u_i(a))^2+\operatorname{Var}(Y_i\mid A=a),
\]
so the noisy target risk upper-bounds the corresponding squared payoff bias.
On $E_{\rm target}$,
\[
|\E_t[u_i-f]|\le\sqrt{r_{k,i,t}}\le\sqrt{u_{k,i,t}}.
\]
Maximizing first over $(i,b)$ and including the no-deviation option yields
\begin{align*}
\Gap_u(q_k)
\le{}&\Gap_{\widehat u_k}(q_k)
+\max_i\sqrt{u_{k,i,q_k}}\\
&+\max_{i,b}\sqrt{u_{k,i,D_{i,b}(q_k)}}=C_k.
\end{align*}
All $q_k$ and deviations appear in the common target union.  The joint
probability claim follows exactly as in Theorem~\ref{thm:oracle}.
\hfill$\square$

This result is deliberately candidate-specific.  Comparing $C_k$ values does
not generally order the true gaps because each certificate contains a
different predicted gap and may have different slack.  Likewise, the
target-risk oracle minimizes a maximum squared prediction risk, not
$\Gap_u(q_k)$.

\section{Proof of Off-Support Non-Identifiability}
\label{app:proof-impossibility}

Let $H\in\{0,1\}$ index the games in
Proposition~\ref{prop:impossibility}, and let $P_h$ be the law of a full logged
dataset under game $h$.  Since $\nu(A_1=0)=1$, both $A_1$ and $A_1A_2$ vanish
almost surely.  Player 1's conditional mean payoff is therefore $0.4$ under
both games on the support of $\nu$.  All other payoffs, action probabilities,
and noise laws coincide; hence $P_0=P_1$ for every sample size.  In contrast,
after replacing player 1's action by one,
\[
\theta_0=0.50,\qquad
\theta_1=0.50+0.2\E[A_2]=0.55.
\]

Let $\widehat H$ be any possibly randomized test based on the log.  Because
$P_0=P_1$, there is a common value
$x=P_0(\widehat H=1)=P_1(\widehat H=1)$.  Its two error probabilities are
$x$ and $1-x$, so
\[
\max_{h\in\{0,1\}}\Pr_h(\widehat H\ne h)
=\max\{x,1-x\}\ge\frac12.
\]
An independent fair coin attains $1/2$, proving the minimax equality.  Since
$\theta_0\ne\theta_1$, the same argument establishes that this replacement
payoff is not point identifiable from the log.

\section{Algorithm and Computational Cost}
\label{app:algorithm}

\begin{algorithm}[ht]
\caption{\sirv{} for finite games}
\label{alg:full-sirv}
\begin{algorithmic}[1]
\Require training log $D_{\rm tr}$, calibration log $D_{\rm cal}$,
known $\nu$, ranks $1{:}K_{\max}$, $\alpha,\tau,\epsilon_{\rm cert},\delta$,
confidence procedure $\mathcal I$
\For{$k=1,\ldots,K_{\max}$}
  \State $\widehat u_k\gets\textsc{FitRank}(D_{\rm tr},k)$
  \State $\sigma_k\gets\textsc{SolveCCE}(\widehat u_k)$
  \State $q_k\gets(1-\alpha)\nu+\alpha\sigma_k$
\EndFor
\State $\T\gets\emptyset$
\For{$j=1,\ldots,K_{\max}$}
  \State $\T\gets\T\cup\{q_j\}$
  \For{$i\in[N]$ and $b\in\A_i$}
    \State add $D_{i,b}(q_j)$ to $\T$
  \EndFor
\EndFor
\If{$\exists t\in\T,a\in\A:t(a)>0$ and $\nu(a)=0$}
  \State \Return \textsc{abstain-unsupported}
\EndIf
\State $\{[\ell_{k,i,t},u_{k,i,t}]\}_{k,i,t}
\gets\mathcal I(D_{\rm cal},\{\widehat u_k\},\T,\delta)$
\If{$\max_{k,i,t}(u_{k,i,t}-\ell_{k,i,t})/2>\epsilon_{\rm cert}$}
  \State \Return \textsc{abstain-imprecise}
\EndIf
\State $L_k\gets\max_{i,t}\ell_{k,i,t}$ and
$U_k\gets\max_{i,t}u_{k,i,t}$ for all $k$
\State \Return $\min\{k:U_k\le\min_jU_j+\tau\}$ and its certificate
\end{algorithmic}
\end{algorithm}

For either profile-based construction, let $P=|\A|$ and $T=|\T|$.  After model
fitting and CCE solving, forming all
calibration losses costs $O(|D_{\rm cal}|K_{\max}N)$ table lookups and
arithmetic operations; integrating the endpoints costs
$O(K_{\max}NPT)$.  The implementation stores all intervals; memory beyond the
fitted models is $O(K_{\max}N(P+T)+PT)$.  In
our three-player binary setting $P=8$, $K_{\max}=3$, and the implementation
has 21 named target slots (at most 21 distinct distributions), so the
validation calculation is negligible relative to data generation and model
fitting.  The procedure requires neither density-ratio estimation nor
iterative validation.

\section{Random-Game Generator and Experimental Grid}
\label{app:generator}

For every player $i$ and order $d$, the generator enumerates subsets
$S\subseteq[N]$ satisfying $|S|=d$ and $i\in S$.  It draws independent signs
$s_{i,S}\in\{-1,1\}$ and magnitudes
$m_{i,S}\sim\operatorname{Unif}[0.5,1.5]$, then sets
\[
\theta_{i,S}=B_d s_{i,S}
\frac{m_{i,S}}{\sum_{T:|T|=d,i\in T}m_{i,T}}.
\]
Thus the absolute coefficients at each player's order sum to $B_d$.  Every
highest-order coefficient is nonzero, so the generating order is exactly
$K$.  We retain every generated game without conditioning on equilibrium
properties or evaluation outcomes.

\begin{table*}[t]
\caption{Experimental designs.  ``Log.'' is the number of logging regimes,
``Noise'' is the number of observation-noise conditions, and the last column
counts configurations except where marked as scoped decisions.}
\label{tab:grid}
\centering
\small
\begin{tabular}{lccccccr}
\toprule
Study & Generating family & Candidates & Games & Log. & $n$ values
& Noise & Rows \\
\midrule
Factorial & $K=1;\ K=2,3,\ h\in\{.04,.12\}$
& $\{1,2,3\}$ & 2,048/family & 4 & 3 & 1 & 245,760 \\
Misspecified & $K=3,\ h=.12$ & $\{1,2\}$ & 2,048 & 3 & 3 & 2 & 73,728 \\
Congestion & affine/quadratic latency & $\{1,2,3\}$
& 256/variant & 3 & 2 & 1 & 6,144 \\
Scaling & normalized congestion, $N\in\{3,5,8\}$ & $\{1,2,3\}$
& 64/stratum & 3 & 3 & 1 & 6,912 \\
Target scope & scaling games at $N=5$ & $\{1,2,3\}$
& 128 & 3 & 1 & 1 & 3,072 decisions \\
\bottomrule
\end{tabular}
\end{table*}

Every cell contains two repetitions.  The factorial study has balanced,
moderate, thin, and unsupported logging.  The misspecified study omits
unsupported logging and crosses the three supported regimes with homoscedastic and
heterogeneous noise.  It therefore contains
$2{,}048\times3\times3\times2\times2=73{,}728$ rows, with 2,048 independent games in
each noise condition.  These are the same 2,048 rank-3, $h=.12$ games as in the
factorial grid, perturbed by candidate-rank restriction and noise condition;
the study is paired rather than an independent holdout.  Under
heterogeneous noise, the half-width table is
drawn once per game and shared across logging regimes, sample sizes, and
repetitions.

Deterministic arithmetic seeds preserve pairing across conditions.  For game
index $g$, let
$b=\lfloor g/64\rfloor$, $r=g\bmod64$, and $o=10{,}000{,}000b$.  Then
\begin{align*}
s_{\rm game}&=30{,}000+1{,}000K+r+o,\\
s_{\rm log}&=60{,}000+10{,}000K+100r
              +\mathrm{regime\_id}+o,\\
s_{\rm data}&=1{,}000{,}000+100{,}000K+1{,}000r+o\\
&\quad+100\,\mathrm{regime\_id}+10\,n_{\rm id}+\mathrm{repetition},
\end{align*}
The heterogeneous half-width seed is $4{,}000{,}000+r+o$.
The data seed excludes $h$, preserving paired actions, observation
noise, and splits across signal levels.  The bootstrap seed is 202,703.

\paragraph{Fitting and CCE solver.}
Each rank uses all Boolean monomials up to that rank, ridge $10^{-8}$, and
post-fit clipping to $[0,1]$.  The CCE linear program maximizes predicted mean
welfare subject to six pure-replacement incentive constraints and the simplex
constraints.  We first use HiGHS with its default settings.  If it reports
failure or its solution does not pass the $10^{-8}$ predicted-CCE-gap check,
we retry the identical linear program with presolve disabled and primal/dual
feasibility tolerances $10^{-9}$.  Every accepted solution passes the same
$10^{-8}$ check.
Solver outputs with absolute mass below $10^{-12}$ are set to zero and
renormalized.  Rank-1 controls can admit multiple welfare optima, so
we treat all outputs as solver-selected rather than unique CCEs.

\section{Congestion-Game Validation}
\label{app:congestion}

We test dependence on the polynomial coefficient generator using weighted
atomic singleton congestion games, a standard resource-allocation
family~\cite{harks2012weighted}.  Each of three players chooses one of two
resources.  Player weights are drawn independently from
$\operatorname{Unif}[.8,1.2]$ and resource load is
$x_r(a)=\sum_j w_j\mathbf 1\{a_j=r\}$.  For each resource,
\[
\ell_r(x)=b_r+s_rx+c_rx^2,
\qquad
u_i(a)=.90-.80w_i\ell_{a_i}(x_{a_i}(a)),
\]
where $b_r\sim\operatorname{Unif}[.05,.12]$ and
$s_r\sim\operatorname{Unif}[.04,.08]$.  The affine variant fixes $c_r=0$;
the quadratic variant draws $c_r\sim\operatorname{Unif}[.006,.014]$.
These ranges keep normalized payoffs away from clipping while producing
nondegenerate load effects over all feasible resource loads; they are not a
canonical benchmark distribution.
All 512 payoff tables lie in $[.25,.90]$.  Direct multilinear expansion of
each table verifies degree 2 for every affine game and degree 3 for every
quadratic game; these labels are computed from the tables rather than assigned
from the generator.  No game is filtered by its equilibrium or evaluation
outcome.
Moreover, mean welfare is $.90-(.80/3)$ times total weighted latency, so
maximizing it is equivalent to the standard congestion-efficiency objective.

For each variant we draw 256 independent games and apply the unchanged seven
selectors, candidate ranks $\{1,2,3\}$, split, thresholds, and noise model.
Balanced, moderate, and thin logging are crossed with
$n\in\{80{,}000,320{,}000\}$ and two repetitions, yielding 6,144 evaluation
runs.  Every randomized component uses a deterministic namespace disjoint
from the factorial experiments.  The inferential unit remains the game;
paired effects are first
averaged across its 12 conditions and then bootstrapped over 512 games.

\begin{figure*}[t]
  \centering
  \includegraphics[width=.98\textwidth]{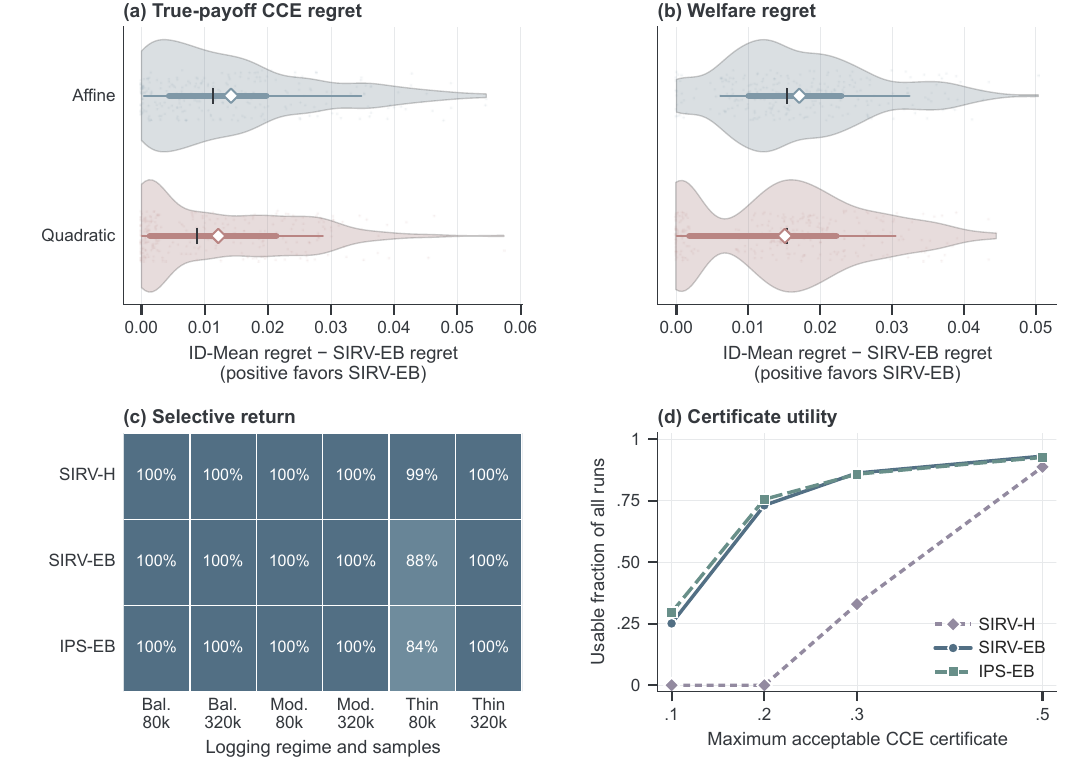}
  \caption{Validation on 512 weighted singleton congestion
  games.  Panels (a--b) show ID-Mean minus SIRV-EB regret under ID fallback,
  so positive values favor SIRV-EB; all 256 paired game effects per latency
  family are shown;
  diamonds are means, vertical ticks are medians, thick segments are IQRs,
  and thin segments span the 10th--90th percentiles.  Panel (c) reports return
  rates over all 6,144 runs.  Panel (d) gives the fraction of evaluation runs
  that both return and meet each CCE-certificate threshold.}
  \Description{Two distribution panels show predominantly positive CCE- and
  welfare-regret effects for affine and quadratic congestion games.  A return
  heatmap is nearly full except under thin logging at 80 thousand samples.
  Certificate-utility curves show empirical-Bernstein methods certifying more
  rows at strict thresholds than the Hoeffding control.}
  \label{fig:congestion}
\end{figure*}

Across both variants, SIRV-EB fallback reduced mean CCE regret by 0.01317
(95\% paired-bootstrap interval $[0.01211,0.01425]$) relative to ID-Mean.
The game-level effect was positive for 93.55\% of games, tied for 6.25\%, and
negative for 0.20\%.  The mean welfare-regret effect was 0.01608
($[0.01514,0.01702]$), with 94.53\% positive, 4.69\% tied, and 0.78\%
negative game means.  Affine and quadratic CCE effects were 0.01417 and
0.01216, respectively, so the combined result is not driven by one latency
family (Figure~\ref{fig:congestion}a--b).

\begin{table}[t]
  \centering
  \caption{All selectors on the congestion panel.  Regrets are game-macro
  means under ID fallback; lower is better.}
  \label{tab:congestion-selectors}
  \small
  \begin{tabular}{@{}lrrr@{}}
    \toprule
    Method & Return (\%) & CCE regret & Welfare regret \\
    \midrule
    ID-Mean      & 100.00 & .01539 & .01887 \\
    ID-Max       & 100.00 & .01433 & .01785 \\
    Target-Point & 100.00 & .00165 & .00198 \\
    Target-IPS   & 100.00 & .00165 & .00197 \\
    IPS-EB       &  97.27 & .00230 & .00290 \\
    SIRV-H       &  99.85 & .00181 & .00225 \\
    SIRV-EB      &  97.97 & .00223 & .00279 \\
    \bottomrule
  \end{tabular}
\end{table}
The point selectors attain the lowest average regrets in this congestion
study;
the interval methods add simultaneous certificates and explicit abstention,
not strategic dominance.

SIRV-EB returned on 6,019/6,144 rows (97.97\%), compared with 6,135 (99.85\%)
for SIRV-H and 5,976 (97.27\%) for IPS-EB.  All three confidence methods
returned in all balanced, moderate, and 320k cells; at 80k under thin logging, the aggregate
rates were 99.1\%, 87.8\%, and 83.6\%, respectively
(Figure~\ref{fig:congestion}c).  On 6,019 common SIRV-H/SIRV-EB returns, the
median CCE certificate fell from 0.34412 to 0.14063, a 59.1\% reduction.
Among all evaluation runs, SIRV-EB both returned and certified thresholds
0.1, 0.2, 0.3, and 0.5 in 25.13\%, 73.16\%, 86.31\%, and 93.18\% of cases
(Figure~\ref{fig:congestion}d).  No covered return violated its target-risk
budget or CCE certificate.  The congestion family remains synthetic, but its
congestion construction is independent of Eq.~\eqref{eq:generator} and
therefore tests whether positive average strategic effects and certificate
behavior transfer beyond that coefficient generator.  It does not test broad
model-class misspecification.

\section{Scaling and Target-Set Analysis}
\label{app:scaling-analysis}

\subsection{Variable-size congestion design}

We next vary the number of players while preserving the finite binary-action
setting of the theory.  For $N\in\{3,5,8\}$, define normalized load
$\widetilde x_r(a)=\sum_jw_j\mathbf 1\{a_j=r\}/\sum_jw_j$ and
\[
\ell_r(\widetilde x)=b_r+s_r\widetilde x+c_r\widetilde x^2,
\qquad
u_i(a)=.90-.80w_i\ell_{a_i}(\widetilde x_{a_i}(a)).
\]
Weights and intercepts use the ranges in
Appendix~\ref{app:congestion}; here
$s_r\sim\operatorname{Unif}[.16,.30]$.  Affine games set $c_r=0$, and
quadratic games draw $c_r\sim\operatorname{Unif}[.08,.16]$.  Normalizing the
load keeps every enumerated payoff in $[0,1]$ as $N$ changes.  Direct
multilinear expansion verifies degree 2 and 3, respectively, for every table.

Each of the six $N$--latency strata contains 64 independent games, giving 384
games.  We use the unchanged candidates, selectors, split, noise,
$\alpha,\tau,\epsilon_{\rm cert}$, and $\delta$.  Balanced, moderate, and thin
logging are crossed with
$n\in\{80{,}000,320{,}000,1{,}280{,}000\}$ and two repetitions, for 6,912
experimental configurations.  Conditions are averaged within game before
inference.

\begin{figure*}[t]
  \centering
  \includegraphics[width=.98\textwidth]{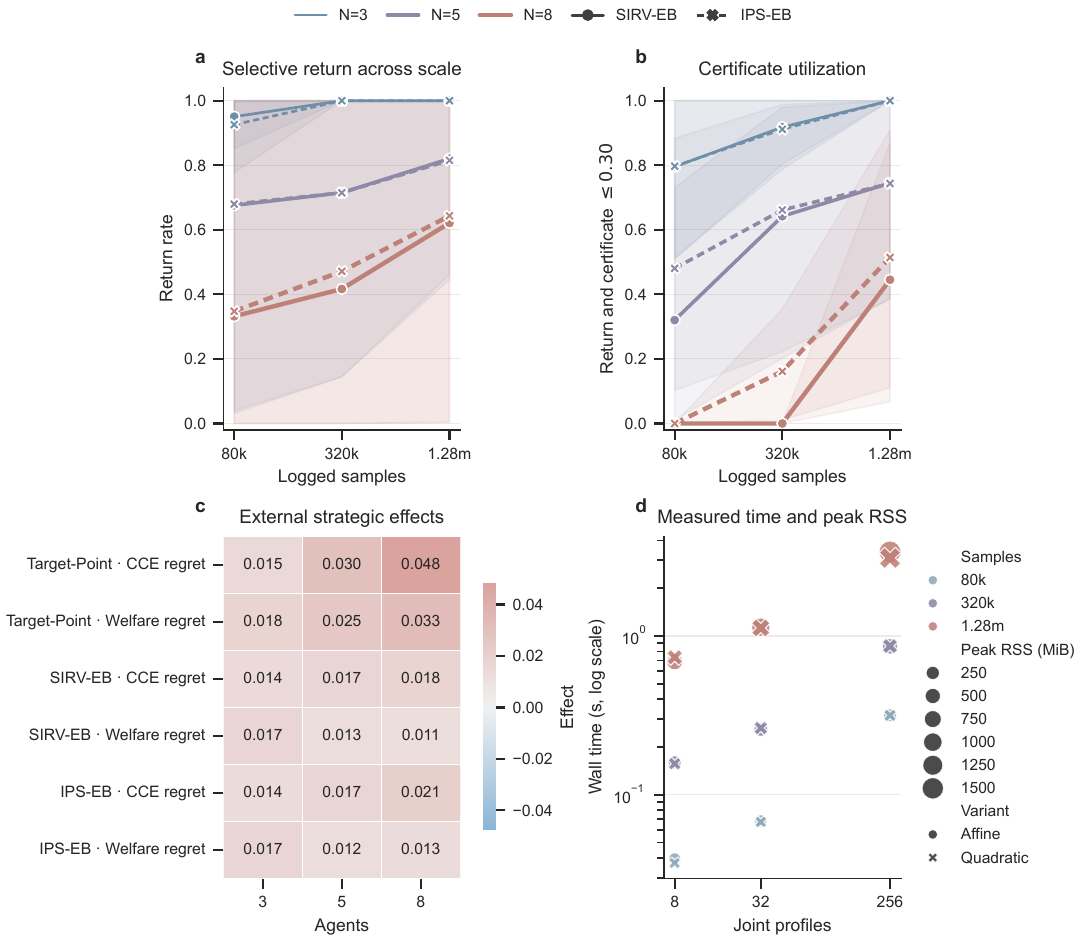}
  \caption{Variable-size congestion analysis.  Panels (a--b) report selective
  return and the fraction of configurations that both return and have a CCE
  certificate at most .30.  In panel (a), points average the six
  latency-family--logging cells and
  bands span their descriptive 10th--90th percentiles.  Panel (b) first
  combines latency families, then summarizes the three logging regimes in the
  same way.  Panel (c) reports
  equal-game mean ID-Mean-minus-method effects, so positive favors the named
  method.  Panel (d) profiles one fixed balanced case per
  $N$--latency--sample combination in a separate single-threaded process;
  color encodes sample size, marker encodes latency family, and area encodes
  peak resident memory.  Time covers the complete shared candidate-fitting,
  selection, and certification computation rather than one selector alone.}
  \Description{Two line panels show return and certificate utilization as
  logged sample size increases for three, five, and eight agents.  A heatmap
  reports CCE- and welfare-regret effects for three methods.  A log-scale
  scatter plot relates end-to-end computation time to the number of joint
  profiles and sample size, with marker area representing peak memory.}
  \label{fig:scaling-analysis}
\end{figure*}

The resource measurements isolate one process per case and are not
inferential replicates.  Each timed call computes all seven selectors from one
shared fit and candidate set; it is not a per-selector SIRV-EB runtime.
Table~\ref{tab:scaling-resources} reports the larger
of the affine and quadratic measurements at $n=1.28$ million.  All 18
resource measurements returned a SIRV-EB choice.  Enumeration remained
practical at $N=8$ in this measured setting, but the joint-profile count is
exponential in $N$ and the measurements do not establish polynomial
scalability.

\begin{table}[t]
  \centering
  \caption{Structure and measured resources at $n=1.28$ million.  Features
  count all Boolean monomials through rank 3; targets count the common union
  for three candidates.  RSS and end-to-end computation time are maxima over the
  two latency families.}
  \label{tab:scaling-resources}
  \small
  \begin{tabular}{@{}rrrrrr@{}}
    \toprule
    $N$ & Profiles & Payoff entries & Features & Targets & RSS / shared time \\
    \midrule
    3 & 8   & 24    & 8  & 21 & 582 MiB / .73 s \\
    5 & 32  & 160   & 26 & 33 & 943 MiB / 1.14 s \\
    8 & 256 & 2,048 & 93 & 51 & 1,567 MiB / 3.38 s \\
    \bottomrule
  \end{tabular}
\end{table}

Across the 2,304 configurations and 128 games at each $N$, SIRV-EB return
fell from 98.35\% at $N=3$ to 73.70\% at $N=5$ and 45.66\% at $N=8$.
The fractions that both returned and had a CCE certificate at most .30 were
90.45\%, 56.90\%, and 14.84\%, respectively.  Mean ID-Mean-minus-SIRV-EB
CCE-regret effects remained positive: .01386
$[.01171,.01608]$, .01672 $[.01374,.01990]$, and .01802
$[.01553,.02053]$.  Thus strategic mean effects did not disappear on this
grid, but selective certificate utility deteriorated as the enumerated game
and target sets grew (Figure~\ref{fig:scaling-analysis}).

\begin{table*}[t]
  \centering
  \caption{Absolute scale-panel outcomes after averaging repeated conditions
  within each of 128 games per $N$.  Regrets use ID fallback and lower is
  better.  Certificate medians condition on return; utilization keeps all
  2,304 configurations per $N$ in the denominator.}
  \label{tab:scaling-methods}
  \small
  \begin{tabular}{@{}rllrrrr@{}}
    \toprule
    $N$ & Method & Return (\%) & CCE regret & Welfare regret & Median cert. & Cert. $\leq .30$ (\%) \\
    \midrule
    3 & ID-Mean      & 100.00 & .0151 & .0184 & --    & -- \\
    3 & Target-Point & 100.00 & .0004 & .0004 & --    & -- \\
    3 & SIRV-EB      &  98.35 & .0012 & .0013 & .1206 & 90.45 \\
    3 & IPS-EB       &  97.53 & .0014 & .0016 & .1063 & 90.28 \\
    \midrule
    5 & ID-Mean      & 100.00 & .0305 & .0250 & --    & -- \\
    5 & Target-Point & 100.00 & .0002 & $<.0001$ & -- & -- \\
    5 & SIRV-EB      &  73.70 & .0137 & .0125 & .1730 & 56.90 \\
    5 & IPS-EB       &  73.65 & .0138 & .0126 & .1404 & 62.85 \\
    \midrule
    8 & ID-Mean      & 100.00 & .0480 & .0328 & --    & -- \\
    8 & Target-Point & 100.00 & .0002 & $<.0001$ & -- & -- \\
    8 & SIRV-EB      &  45.66 & .0300 & .0216 & .4523 & 14.84 \\
    8 & IPS-EB       &  48.74 & .0269 & .0196 & .3022 & 22.53 \\
    \bottomrule
  \end{tabular}
\end{table*}

\begin{table}[t]
  \centering
  \caption{Smallest tested $n$ at which SIRV-EB return is at least .90 and
  its median returned CCE certificate is at most .30.  ``Not met'' means the
  joint criterion failed at every tested $n$ through 1.28 million.}
  \label{tab:scaling-grid}
  \small
  \begin{tabular}{@{}rlll@{}}
    \toprule
    $N$ & Balanced & Moderate & Thin \\
    \midrule
    3 & 80k   & 80k   & 320k \\
    5 & 80k   & 320k  & Not met \\
    8 & 1.28m & Not met & Not met \\
    \bottomrule
  \end{tabular}
\end{table}

Table~\ref{tab:scaling-grid} is a descriptive diagnostic over three tested
sample sizes, not a sample-complexity estimate or an extrapolation beyond the
grid.  Target-Point always returned and had the lowest absolute regret in
this family; the interval methods add simultaneous guarantees and abstention,
not strategic dominance (Table~\ref{tab:scaling-methods}).

\subsection{Target scope and heterogeneous-noise ordering}

The target-set ablation reuses the same fitted candidates, splits, and
observations for four decisions on the 128 independent five-agent games at
$n=320{,}000$.  The original \emph{common+deviation} design compares every
candidate on the union of all deployments and unilateral replacements.
\emph{Local+deviation} evaluates each candidate only on its own deployment
and replacements, making support and precision eligibility candidate-local.
The two deployment-only variants omit all replacement targets.  Thus local
designs change the comparator, while deployment-only designs cannot produce
the CCE certificate in Theorem~\ref{thm:cce}.

\begin{figure*}[t]
  \centering
  \includegraphics[width=.98\textwidth]{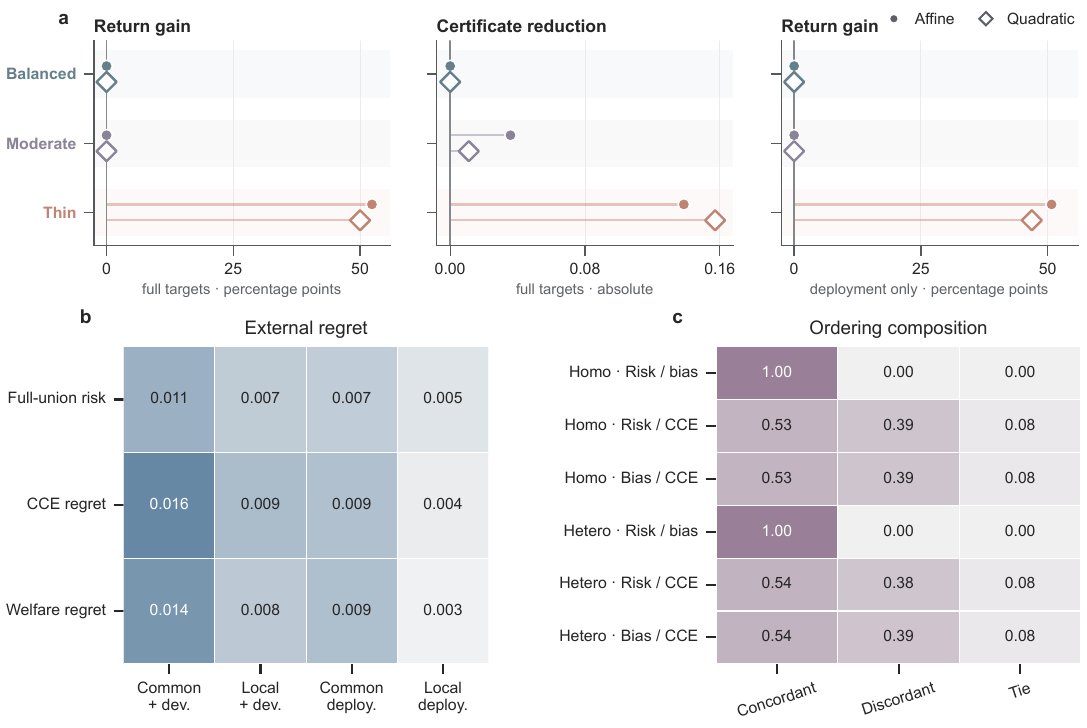}
  \caption{Target-set and noise-ordering analysis.  Panel (a) gives paired
  local-minus-common effects in six prespecified 128-configuration cells.
  Horizontal segments begin at zero; circles denote affine games, diamonds
  denote quadratic games, and row color denotes logging regime.  The first
  two facets show the full-target return-rate gain and reduction in the median
  returned CCE certificate; the third shows the deployment-only return-rate
  gain, for which no deviation certificate is available.  Positive values
  favor the local design on the displayed quantity.  Panel (b) evaluates all
  four designs using full-common exact risk and true-payoff regrets under ID
  fallback; lower is better.  Panel (c) gives pairwise ordering composition
  among noisy target risk, noise-free payoff bias, and true CCE gap over 2,048
  games per noise condition.}
  \Description{A three-facet contrast forest reports local-minus-common
  changes in return rate and CCE-certificate magnitude for affine and
  quadratic games under balanced, moderate, and thin logging.  A heatmap
  reports three external regrets for the four target designs.  A second
  heatmap shows concordant, discordant, and tied ordering fractions under
  homoscedastic and heterogeneous noise.}
  \label{fig:target-noise-analysis}
\end{figure*}

Common+deviation returned on 71.48\% of the 768 configurations.  Making the
full target set candidate-local raised return to 88.54\%; its median returned
CCE certificate was .16662 versus .16018 for the common design.  Common and
local deployment-only return rates were 82.55\% and 98.83\%, respectively,
but those decisions retain only the deployment-target risk bound and do not
certify deviation incentives.  All three logging regimes have positive
support, so the return differences in this panel arise from precision scope,
not recovered support.  Descriptively, each local design also had lower mean
full-common risk, CCE, and welfare regret than its corresponding common design
(Figure~\ref{fig:target-noise-analysis}b).  This is a performance--
comparability tradeoff, not dominance over the common construction: the
local rule no longer asks every candidate to meet one shared target criterion.
Among simulated rows on the scoped simultaneous interval event, no returned
decision violated its scoped risk bound or candidate CCE certificate.

Finally, we quantify how heterogeneous observation variance changes the
two-candidate ordering in the rank-misspecified study.  Deltas are computed
for rank 2 minus rank 1 after averaging the 18 experimental configurations
(three logging regimes, three sample sizes, and two repetitions) within each
game and noise condition.  Noisy target risk and noise-free squared
payoff bias agreed on all 2,048 homoscedastic games and on 2,042/2,048
heterogeneous games (99.71\%); the paired change was $-0.29$ percentage
points (95\% game-bootstrap interval $[-0.54,-0.10]$).  In contrast, among games non-tied on both
quantities, strict risk--CCE ordering agreement was 1,085/1,888 (57.47\%) and
1,101/1,889 (58.28\%), respectively; about 7.8\% of all comparisons were tied
in at least one CCE-related quantity.  Heteroscedasticity therefore
rarely changed the target-risk versus bias ordering in this perturbation, but
neither ordering was a reliable surrogate for the strategic CCE ordering in
this study.

\section{Additional Experimental Results}
\label{app:additional}

\subsection{Selector definitions and denominators}

For a selective method, ``conditional'' quantities average only its returned
rows.  ``ID fallback'' replaces an abstention with the ID-Mean selected rank
and therefore retains all supported rows.  The main external comparisons use
ID fallback; conditional means are not compared across methods because their
returned subsets differ.  Point comparators return on every row except
Target-IPS, which
returns on all supported rows and refuses all unsupported rows.  The support
rule is thus not shared by all seven methods.

\subsection{Target-risk alignment}

\begin{figure*}[t]
  \centering
  \includegraphics[width=.97\textwidth]{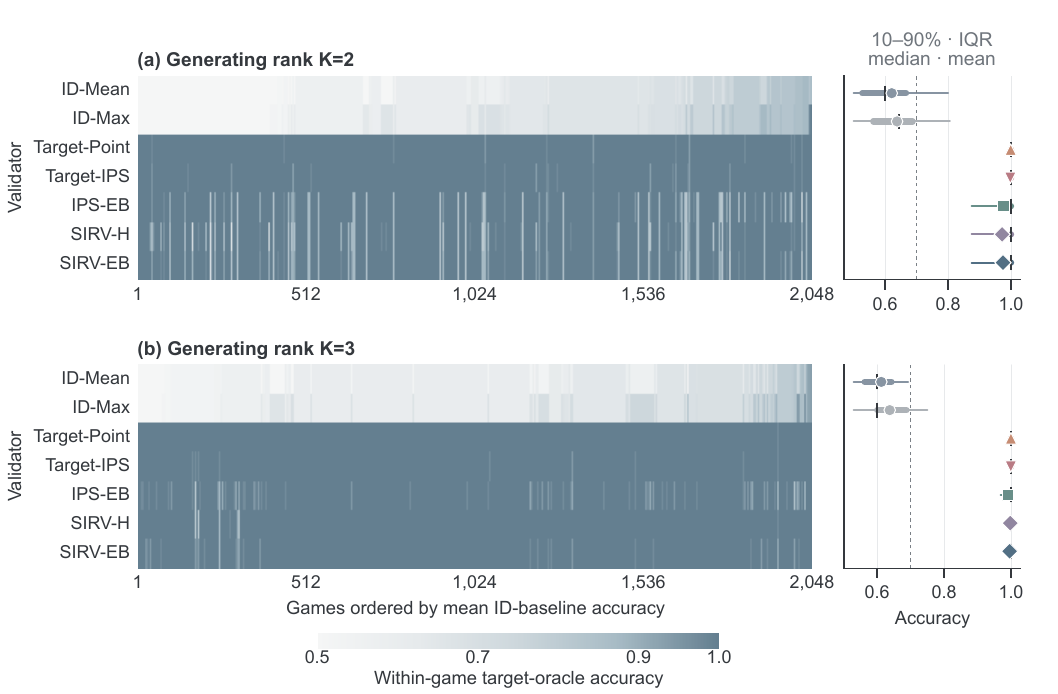}
  \caption{Target-risk-oracle accuracy on common supported returns.
  Rows share 2,048 games; marginals show the 10th--90th percentiles, IQR, median,
  and mean.  This measures $R_k$-order recovery, not strategic value.}
  \Description{Two seven-by-2,048 heat strips compare ID-Mean, ID-Max,
  Target-Point, Target-IPS, IPS-EB, SIRV-H, and SIRV-EB on exact
  target-risk-oracle accuracy for generating ranks two and three.  Marginal
  interval plots summarize variation across games.  Target-aware rows are
  generally more accurate than in-distribution rows.}
  \label{fig:accuracy}
\end{figure*}

At $\tau=0.002$, Target-Point had game-macro accuracy 0.9989 and
0.9998 for generating ranks 2 and 3, respectively, on the common returned
set.  Target-IPS had accuracy 0.9978 and 0.9992, while SIRV-EB had 0.9745 and
0.9958.
These values show that the estimators track their declared target-risk
criterion.  They cannot establish a reduction in true CCE or welfare regret,
which is why Figures~\ref{fig:external}--\ref{fig:misspecified} use the payoff-table
outcomes instead.

\begin{figure*}[t]
  \centering
  \includegraphics[width=.86\textwidth]{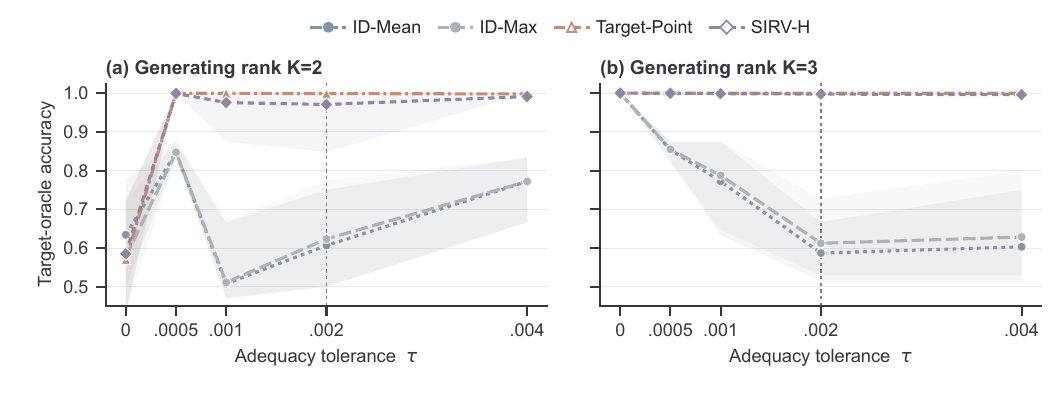}
  \caption{Target-risk-oracle accuracy across
  $\tau\in\{0,.0005,.001,.002,.004\}$ on common supported returns.  Points are
  equal-game means and bands show the descriptive 10th--90th percentile range
  across 2,048 games.  The vertical marker denotes the main value
  $\tau=.002$.  Accuracy is not monotone in $\tau$ because the smallest
  adequate oracle rank changes with the tolerance.}
  \Description{Two panels show target-risk-oracle accuracy versus adequacy
  tolerance for generating ranks two and three.  Four validators have point
  estimates and 10th-to-90th-percentile bands; a vertical line marks tau equal
  to .002.}
  \label{fig:tau}
\end{figure*}

\subsection{Selective return and certificate thresholds}

For SIRV-EB, balanced and moderate logging returned in every factorial cell.
Thin-logging return rose from 40.94\% at $n=20{,}000$ to 84.27\% at
$n=80{,}000$ and 100\% at $n=320{,}000$.  All 61,440 unsupported rows were
refused.  SIRV-H and IPS-EB also refused every unsupported row; their
supported return rates were 93.05\% and 90.39\%.

Here $B_R=\tau+\omega$ for the profile-based confidence procedures and the corresponding
simultaneous risk budget for IPS-EB.  Among returned rows, the 10th--90th
percentiles of $C_{\widehat k}$ were $[.31872,.68205]$ for SIRV-H,
$[.14683,.50407]$ for SIRV-EB, and $[.13666,.47127]$ for IPS-EB.
The 90th percentiles of realized risk excess divided by its budget were
.0388, .1612, and .1892, respectively; all three medians were zero.  Tighter
budgets therefore improve utilization but need not reduce the realized-to-
allowed ratio.

No covered return in either experimental grid exceeded its target-risk budget
or CCE certificate.  This is a numerical validation over the simulated rows,
not an estimate of zero population failure probability.  In particular, no
confidence procedure certified $C_{\widehat k}\le.05$ in the factorial study.  The
joint-probability theorems and the empirical threshold utilization answer
different questions.

\subsection{External effects and selection stability}

\begin{table*}[t]
\caption{Mean ID-fallback effect relative to ID-Mean, with 95\% paired
base-game bootstrap intervals in brackets.  Positive values indicate lower
regret.  All generating-rank-2 entries are exact zero; misspecified columns use heterogeneous
noise and define regret relative to the better candidate in $\{1,2\}$, not to
the excluded rank-3 deployment.}
\label{tab:external-effects}
\centering
\small
\renewcommand{\arraystretch}{0.82}
\begin{tabular}{lccccc}
\toprule
Method & Rank-2 CCE/W & Rank-3 CCE & Rank-3 welfare
& Misspecified CCE & Misspecified welfare \\
\midrule
ID-Max
& .00000/.00000
& .00042 [.00026,.00058] & .00042 [.00029,.00055]
& .000038 [-.000004,.000117] & .000027 [-.000005,.000085] \\
Target-Point
& .00000/.00000
& .00993 [.00931,.01054] & .00635 [.00588,.00684]
& .01741 [.01569,.01915] & .00449 [.00339,.00561] \\
Target-IPS
& .00000/.00000
& .00993 [.00931,.01054] & .00635 [.00588,.00684]
& .01746 [.01576,.01917] & .00458 [.00349,.00569] \\
IPS-EB
& .00000/.00000
& .00380 [.00357,.00403] & .00244 [.00224,.00265]
& .01246 [.01145,.01351] & .00370 [.00301,.00439] \\
SIRV-H
& .00000/.00000
& .00538 [.00505,.00571] & .00346 [.00320,.00373]
& .01395 [.01277,.01516] & .00408 [.00328,.00486] \\
SIRV-EB
& .00000/.00000
& .00518 [.00486,.00550] & .00328 [.00303,.00355]
& .01362 [.01248,.01479] & .00397 [.00321,.00474] \\
\bottomrule
\end{tabular}
\end{table*}

The positive rank-3 and misspecified-study means are not pointwise dominance
statements.  Under heterogeneous noise, for example, SIRV-EB's game effects
on CCE regret were 49.2\% positive, 25.0\% tied, and 25.9\% negative; its
welfare effects were 47.1\% positive, 20.8\% tied, and 32.1\% negative.
The SIRV-EB mean welfare interval in Table~\ref{tab:external-effects} is
positive despite this negative game-level tail.
The point selectors have larger mean CCE effects than the selective methods
in this family, while the interval methods add simultaneous bounds and
explicit refusal.

\section{Evaluation and Clustered Inference}
\label{app:statistical-analysis}

Exact payoff tables generate the synthetic rewards but are unavailable to all
seven selectors.  They enter only evaluation quantities: exact target risks,
target-oracle labels, true CCE gaps, and true welfare.
The external analyses include abstentions through ID fallback and retain
exact ties and negative effects.

Repetitions, sample sizes, signal levels, and logging regimes are repeated
conditions within a base game rather than additional independent games.  In
the factorial and misspecified studies, paired ID-Mean-minus-method effects
are averaged within game and the 2,048 equally weighted game effects are
resampled within each reported family or noise condition.  The congestion
validation applies the same procedure to 512 games after averaging its 12
conditions per game.  The scaling analysis first averages 18 conditions per
game, then resamples 64 games for a latency-family result or 128 games for a
result combined across both latency families at a fixed $N$.  Every reported
bootstrap interval uses 5,000 resamples with seed 202,703.

\section{Reproducibility Details}
\label{app:reproduction}

All simulations use deterministic pseudorandom seeds.  The main studies
comprise 10,240 payoff tables, 24,576 logging distributions, 245,760 factorial
configurations, and 73,728 misspecified configurations.  The congestion
validation adds 512 payoff tables, 1,536 game--regime assignments, and 6,144
configurations.  The scaling study adds 384 payoff tables, 1,152
game--regime assignments, and 6,912 configurations.  Its target-set ablation
re-evaluates 768 of those configurations under four target scopes, yielding
3,072 decisions, and its resource panel comprises 18 isolated process
measurements.  The noise-ordering analysis is derived from the existing
misspecified configurations and introduces no additional simulated rows.
Statistical analyses treat the base game as the inferential unit.  Experiments
were run with Python 3.13.1, NumPy 1.26.4, pandas 2.2.3, and SciPy 1.15.2.
The isolated resource measurements used one process and one numerical-library
thread.

\section{Scope and Limitations}
\label{app:limitations}

The theory assumes a finite action table and known logging distribution;
estimated propensities, continuous actions, and sequential games require new
error control.  Global-union support and precision abstention are conservative
and may reject a useful supported-subset answer; that restriction changes the
target risk and oracle.  Conversely, positive support may be statistically
negligible, motivating a separate precision check.

Because the target risk in Eq.~\eqref{eq:targetrisk} uses noisy squared loss,
it bounds squared mean-payoff error but also contains conditional reward
variance.  Heterogeneous noise can therefore alter the target-risk oracle.
The robustness study examines one profile-specific variance pattern, not a
general noise-shift benchmark.  The finite-profile construction enumerates
$|\A|=2^N$ profiles in the binary-action panels.  The measurements through
$N=8$ therefore do not establish scalability to large joint-action spaces.

Our evaluation remains synthetic rather than a benchmark of real-world
strategic systems.  The factorial family contains three binary-action players,
linear monomial payoff classes aligned with the fitted hierarchy, bounded
uniform noise, and predicted CCE candidates chosen for welfare.  Its 245,760
rows provide broad paired coverage, not evidence across 245,760 independent
environments; there are 2,048 independent base games per generating order.
The misspecified study reuses the rank-3, $h=.12$ games as a paired
perturbation, and its candidate regrets compare ranks $\{1,2\}$ only.  The
separately generated congestion games reduce dependence on the monomial
coefficient generator.  The scaling panel varies the number of players through
$N=8$, but every game still has two actions per player and retains the same
observation model and candidate solver.  We use base games as inference units
throughout.

Variance-adaptive bounds tighten certificates, but the point selectors have
larger mean CCE-regret reductions in the studied rank-3 conditions.  The
interval methods instead provide simultaneous bounds and abstention.
Heterogeneous-noise outcomes retain a negative game-level tail, and few
SIRV-EB returns meet joint risk-budget .05 and CCE .2 thresholds.  Real logged
strategic interactions are needed to assess external generality.

\end{document}